\documentclass[10pt]{article}
\usepackage[OE]{express2}
\usepackage{titlesec}

\begin{document}

\title{Theory of plasmon reflection by a 1D junction}

\author{Bor-Yuan Jiang,\authormark{1} Eugene J. Mele,\authormark{2} and Michael M. Fogler\authormark{1,*}}

\address{\authormark{1}Department of Physics, University of California San Diego, La Jolla, California, 92093, USA\\
\authormark{2}Department of Physics \& Astronomy, University of Pennsylvania, Philadelphia, Pennsylvania 19104, USA}

\email{\authormark{*}mfogler@ucsd.edu} 



\begin{abstract}
We present a comprehensive study of the reflection of normally incident plasmon waves from a low-conductivity 1D junction in a 2D conductive sheet. 
Rigorous analytical results are derived in the limits of wide and narrow junctions.
Two types of phenomena determine the reflectance, the cavity resonances within the junction and the capacitive coupling between the leads.
The resonances give rise to alternating strong and weak reflection but are vulnerable to plasmonic damping.
The capacitive coupling, which is immune to damping, induces a near perfect plasmon reflection in junctions narrower than $1/10$ of the plasmon wavelength.
Our results are important for
infrared 2D plasmonic circuits utilizing slot antennas, split gates or nanowire gates. They are also relevant for the implementation of nanoscale
terahertz detectors,
where optimal light absorption coincides with the maximal junction reflectance.
\end{abstract}

\ocis{(240.6680) Surface plasmons; (250.5403) Plasmonics; 
	(040.2235) Far infrared or terahertz; (250.6715) Switching.} 


\section{Introduction}

Plasmonics aims to combine the advantages of nanometer scale electronics with the high operating frequency (terahertz and beyond) of photonics \cite{Ozbay2006pmp}.
A promising platform for plasmonics is graphene, which features  high confinement, wide range of operating frequencies $\omega$, long lifetimes, and tunability \cite{Grigorenko2012gp,Fei2012gtg}.
Recent experiments demonstrated long plasmon propagation distance and high quality factor for graphene \cite{Woessner2014hcl, Ni2016uos}, making 2D plasmonic circuitry feasible.
An important basic element in a plasmonic circuit is a switch that has a small size and a large on-off ratio.
Numerous numerical studies of the interaction of plasmons with 1D obstacles has been done in search of such a device \cite{Ryzhii2004pos, Popov2005rtr, Satou2005pot, Huerkamp2011ghe, Popov2012erp, Garcia-Pomar2013sgp, Polanco2013ssp, Woessner2017etp}.
It has been shown that 
a narrow 1D junction of low conductivity in an otherwise uniform 2D conductive sheet can potentially serve as a plasmonic switch \cite{Gomez-Diaz2013gbp, Garcia-Pomar2013sgp}.
This is a surprising result as deeply subwavelength obstacles usually cannot impede the propagation of a wave.
In this work we explain the physical principles
behind the plasmonic interaction with this type of inhomogeneities and
provide analytical solutions for the reflection coefficient, which can be simply understood in terms of  equivalent circuits.
The anomalous complete reflection is revealed to have the same origin as
the stop-band behavior exploited in LC-loaded radio frequency waveguides or various metamaterial structures.



\begin{figure}[t]
	\centering
	\includegraphics[width=0.4\linewidth]{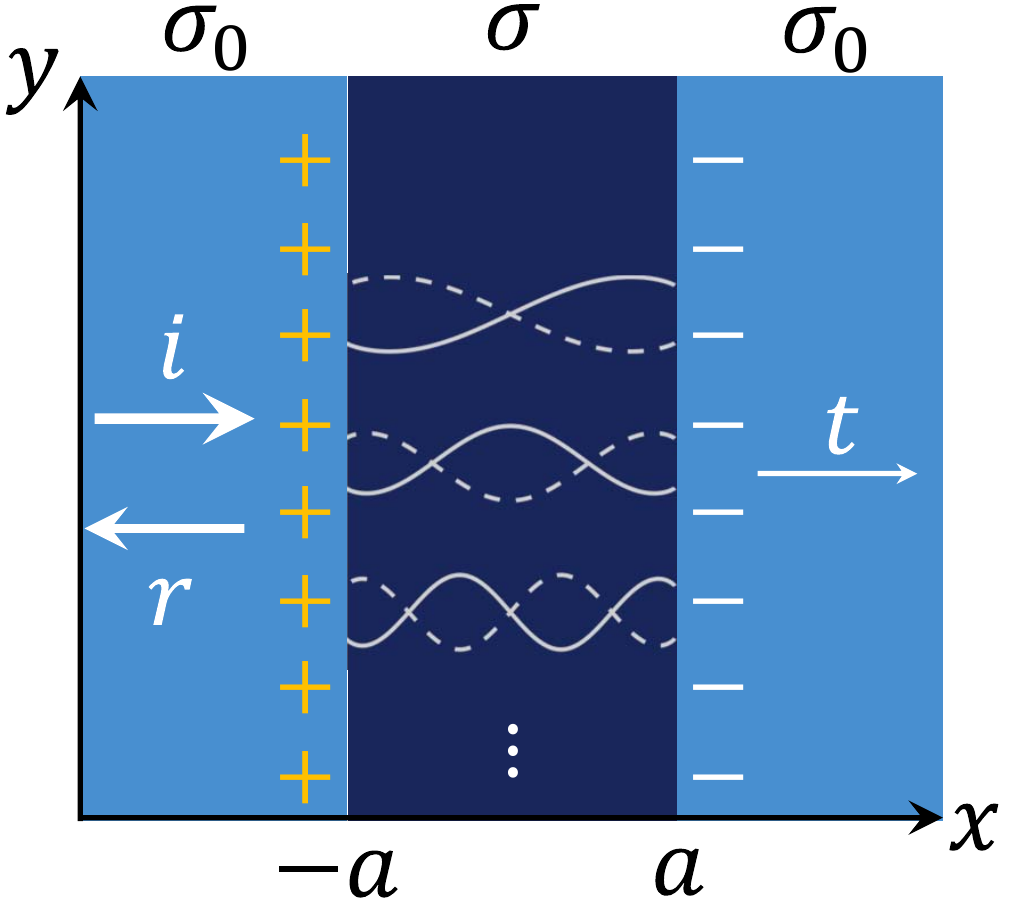}
	\caption{
		A normally incident plasmon is partially reflected and transmitted by a 1D junction of conductivity $\sigma$ different from the background value $\sigma_0$.
		The strength of reflection is determined by two types of effects, 
		the capacitive coupling of the two edges of the leads (represented by the $+$ and $-$ electric
		charges) and the cavity resonances in the strip.
		The field profiles of the first few resonant modes (white solid and dashed
		curves) are calculated for the case of a narrow junction
		with an infinite conductivity contrast between the gap and the leads.
	}
	\label{fig1}
\end{figure}

Plasmons propagating in a uniform conducting film or a two-dimensional electron gas (2DEG) have a momentum $q$
that is inversely proportional to the (frequency-dependent) sheet conductivity $\sigma$, $q=\frac{i\kappa\omega}{2\pi\sigma}$,
where $\kappa$ is the dielectric function of the environment  exterior to the sheet.
A local variation in the conductivity causes a change in $q$ and thus acts as a scatterer for plasmons.
Controlled conductivity variations can be realized in graphene which has a conductivity $\sigma$ determined by its chemical potential.
Using patterned electric gates, the chemical potential can be tuned locally.
We consider 
an idealized case where the sheet conductivity has a piecewise constant 1D profile with a value $\sigma$ inside a strip of width $2a$ and another value $\sigma_0$ in the semi-infinite leads on both sides 
(Fig.~\ref{fig1}).
In practice, the width of the junction is determined by the geometry of the gate and can be as narrow as
a few nanometers for a nanowire or nanotube gate \cite{Jiang2015erg, Jiang2016tpr}. 
Two types of phenomena govern the propagation of plasmons across such a junction, the cavity resonances inside the junction and the capacitive coupling between the leads.
Depending on how the width $(2a)$ of the junction compares to the plasmon wavelength in the leads $(\lambda_0=2\pi/\mathrm{Re}\, q_0)$ and in the junction $(\lambda)$, the strength of these phenomena varies.
For wide junctions $\lambda_0\ll a$, the reflectivity is determined chiefly by the cavity resonances  while the capacitive coupling is negligible.
The resonances give rise to alternating maxima and minima in the reflectance, which are, however, quickly suppressed by plasmonic damping.
For narrow junctions $\lambda_0\gg a$, the capacitive coupling dominates.
This regime can be further divided into two, depending on how $a$ compares to $\lambda$.
When $a \gg \lambda$, the cavity resonances are present but have their amplitudes modified by the capacitive coupling.
When $a \ll \lambda$, instead of the oscillating resonant fields, the junction has a constant electric field like a capacitor and becomes a parallel $LC$ circuit in the dc limit.
A near perfect plasmon
reflection occurs  at a specific conductivity $\sigma$ where the impedance of the $LC$ circuit diverges.
As there is no wave propagation in the junction in this limit, the anomalously strong reflection is robust against plasmonic damping.



\begin{figure}[t]
	\centering
	\includegraphics[width=0.6\linewidth]{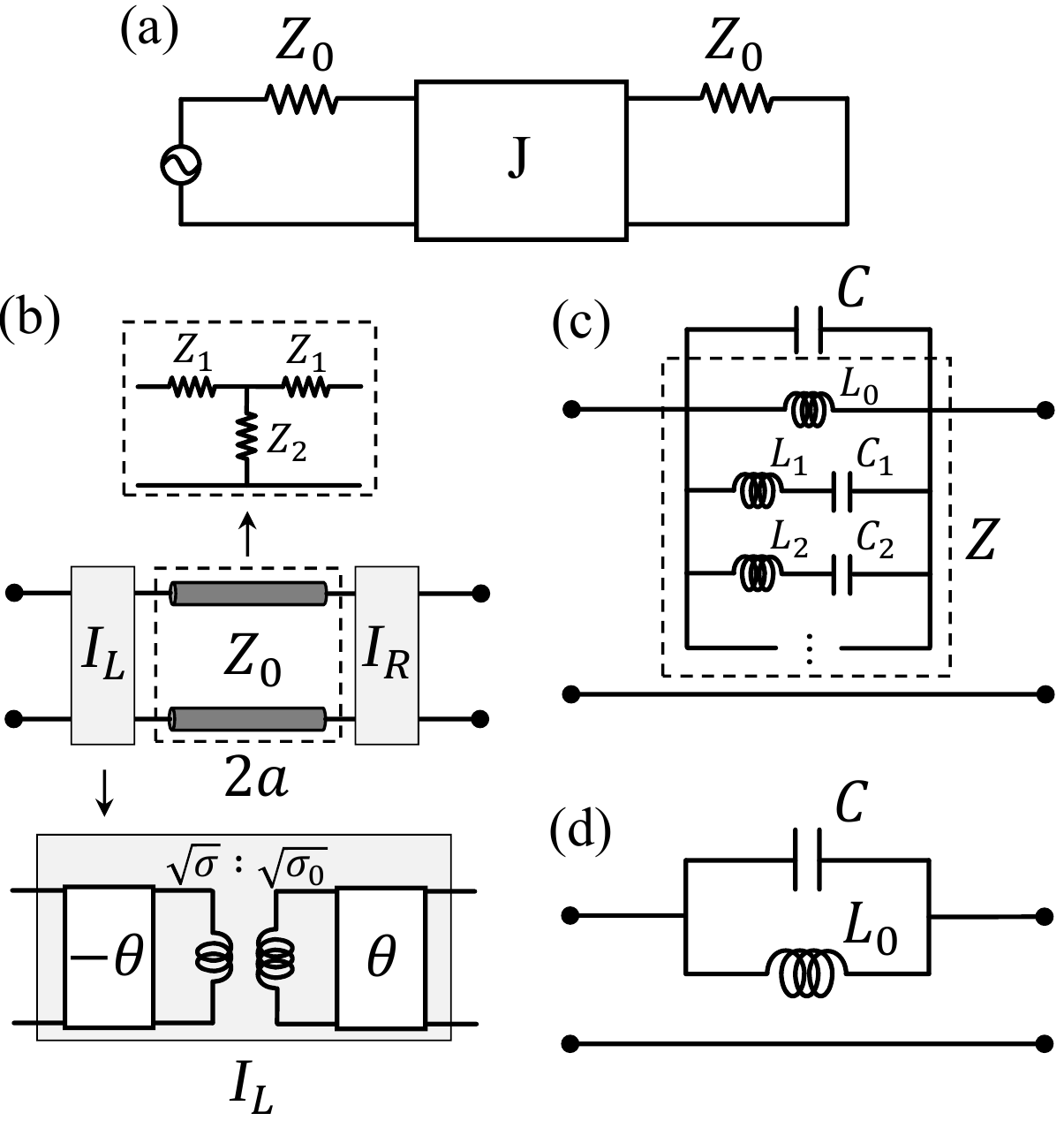}
	\caption{
		(a) Equivalent circuit for the system. The leads have impedance $Z_0$, while the junction $\textrm{J}$ has different representations depending on its width $a$ and conductivity $\sigma$.
		(b) For wide junctions $\textrm{J}$ is represented by two interfaces and a waveguide of length $2a$. The waveguide can be replaced by a $T$-junction, while the interface $I$ consists of two phase shifters and an ideal transformer. 
		The right interface $I_R$ has reversed number of coils and signs of $\theta$ compared to $I_L$.
		(c) For narrow junctions $\textrm{J}$ is a capacitor in parallel with a $LC$ network that describes the cavity resonances.
		(d) In the dc limit, $a\ll\lambda$, the $LC$ network reduces to a single inductance $L_0$.
	}
	\label{fig2}
\end{figure}


\section{Results}
\subsection{Wide junctions} 
\begin{figure}[t]
	\centering
	\includegraphics[width=.7\linewidth]{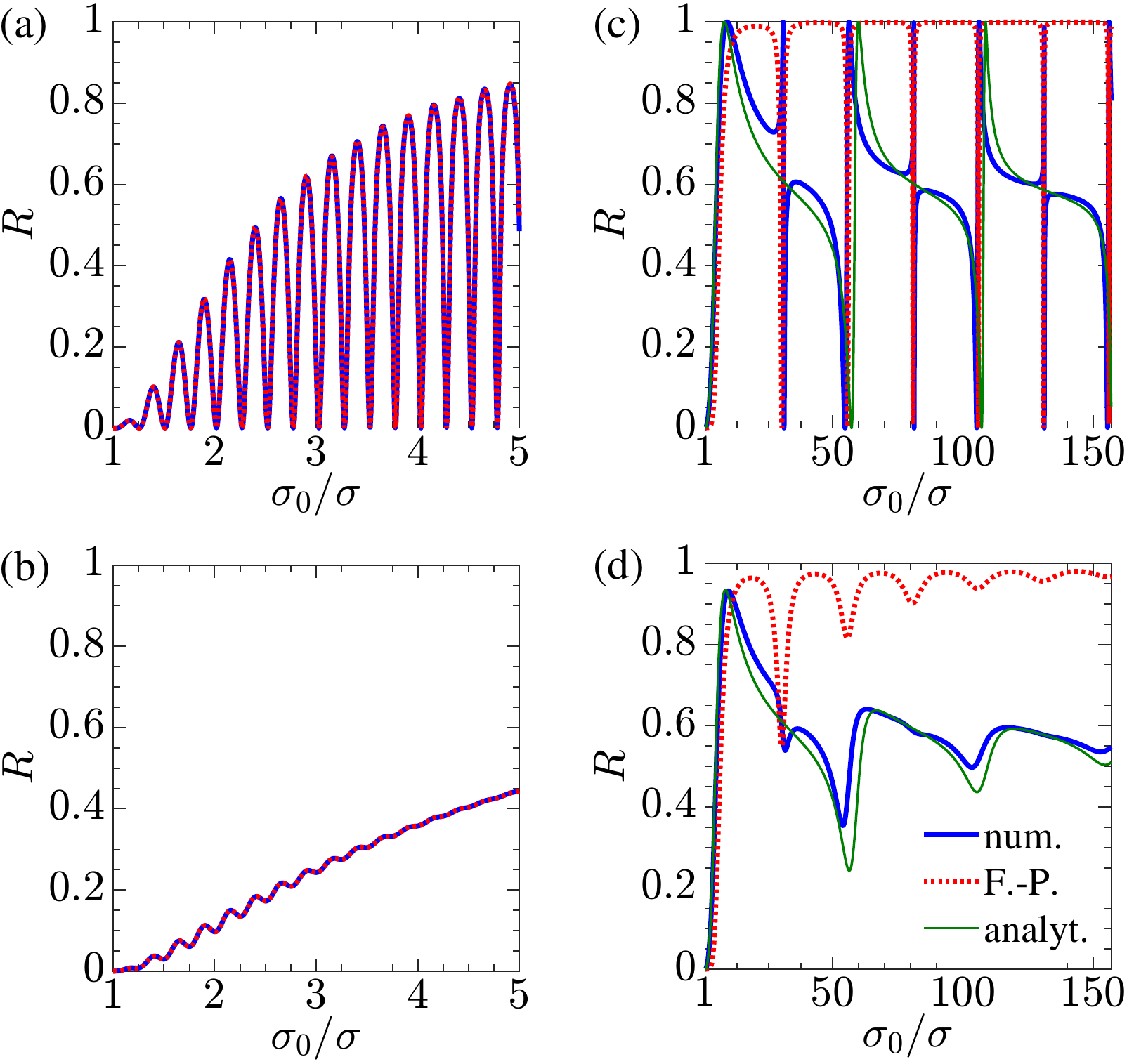}
	\caption{
		(a) Reflectance of a ``wide'' junction, $a=\lambda_0$ without damping.
		(b) Similar quantities  for damping $\gamma=0.05$.
		(c) Reflectance of a ``narrow'' junction, $a=0.01\lambda_0$ without damping. 
		Blue curves are numerical results, dotted red curves are from the F-P formula Eq.~\eqref{eqn:r_t_FP}, and green curves are  from Eq.~\eqref{eqn:r_analyt}.
		(d) Similar quantities for $\gamma=0.05$.
	}
	\label{fig3}
\end{figure}

A wide junction contains two interfaces that partially reflect and transmit plasmons but otherwise do not interact with each other, i.e., it is a plasmonic cavity.
Using the analytical Wiener-Hopf method, the reflection coefficient $r$ for a normally incident plasmon wave from the left at the left ($L$) and the right ($R$) interfaces was found in Ref.~\citenum{Rejaei2015ssp} to be
\begin{equation}
	r_L=e^{i\theta}\frac{\sigma_0-\sigma}{\sigma_0+\sigma}\,,\quad
	r_R=e^{-i\theta}\frac{\sigma-\sigma_0}{\sigma+\sigma_0}\,,\quad
	t_L=t_R=\frac{2\sqrt{\sigma_0\sigma}}{\sigma_0+\sigma}\,.
	\label{eqn:r_t_inter}
\end{equation}
The nontrivial reflection phase shift, frequently overlooked in previous studies \cite{Dyer2012iit,Dyer2013itc}, is given by
\begin{equation}
	\theta=\frac{\pi}{4}-\frac2\pi\int_0^\infty du\,\frac{\tan^{-1}( \frac{\sigma}{\sigma_0}u)}{u^2+1}\,,
\end{equation} 
which approaches $\pi/4$ in the limit $\sigma \to 0$, as in the case where the lead-junction boundary corresponds to a physical termination of the lead \cite{Nikitin2014arp,Kang2017ghs}.
We define the reflection coefficient $r$ to be the prefactor in the asymptotic form  of the scattered potential $\phi_s\simeq re^{-iq_0x}$ at large negative $x$ for an incident plasmon potential $\phi_0=e^{iq_0x}$.
(Our definition differs in the overall sign from Ref.\citenum{Rejaei2015ssp} where $r$ is the reflection coefficient for the current.)
The total reflection coefficient from the left interface of the junction is then found from the usual Fabry-P\'{e}rot (F-P) formula to be 
\begin{equation}
	r_{\mathrm{FP}}^{-1}=\frac{\sigma_0^2+\sigma^2}{\sigma_0^2-\sigma^2}+i\frac{2\sigma_0\sigma}{\sigma_0^2-\sigma^2}\cot(\phi-\theta)\,,
	\label{eqn:r_t_FP}
\end{equation}
where $\phi=2qa$ is the phase accumulated across the junction.
Equation \eqref{eqn:r_t_FP} yields a reflectance that has local minima at resonances, $\phi=n\pi+\theta$, and local maxima at anti-resonances, $\phi=(n+\frac12)\pi+\theta$, where $n$ is a positive integer.
These alternating maxima and minima can be seen in Fig.~\ref{fig3}a, where an example of the reflectance $R=|r|^2$ for a lossless junction of width $a=\lambda_0$ is shown.
Here we have parametrized the plasmon momentum as $q=\frac{2\pi}{\lambda}(1+i\gamma)$ with the dimensionless damping factor defined as $\gamma=\mathrm{Im}\,q/\mathrm{Re}\,q$.
Under these circumstances, the junction acts as a plasmonic switch with a high on-off ratio, tuned by the conductivity contrast $\sigma_0/\sigma$
(or alternatively, by varying the width $a$).
However, the resonances are quickly suppressed by plasmonic damping as shown in Fig.~\ref{fig3}b, removing the switching behavior and rendering wide junctions less desirable for nanoplasmonic applications.
Note that we assume the same damping in the leads and the junction, $\gamma_0=\gamma$, in this work, so that the conductivity ratio $\sigma_0/\sigma$ is real. 
A higher damping in the junction than in the leads further diminishes the resonances but does not alter the results qualitatively 
(see Appendix \ref{sec:damping}.)
Note also that we neglect the screening of plasmons by the gates
forming the junction. This effect is quite small for graphene plasmonic
structures operating in mid-IR with typical parameters $\lambda_0 \sim 100\, \mathrm{nm}$,
$d \sim 300\, \mathrm{nm}$, where $d$ is the distance between the 2D sheet and the gate.
The screening would become important, however, if the plasmon confinement length
$\lambda_0 / 2\pi$ exceeds $d$,
which is the case in the THz domain. In the first approximation,
the screening effectively replaces the long-range Coulomb interaction by
a short-range one, so that the propagation of the plasmons is 
described by a wave equation with a position-dependent $\lambda_0$.
In such a limit, the theoretical problem becomes trivial.
Both the reflection phase shift and
the capacitive coupling of the two sides of the junction can be neglected,
so that the simplest version of the Fabry-Perot formula (with $\theta = 0$) \cite{Dyer2012iit,Dyer2013itc} applies.

A wide junction can be described with an equivalent  circuit. The reflection and transmission coefficients described in Eq.~\eqref{eqn:r_t_inter} is what one would expect at the interface of two waveguides with impedances $\sigma_0$ and $\sigma$ (up to the phase factor in $r$) \cite{Montgomery1948pmc}.
However, the plasmonic wave impedance of a 2DEG, $Z_0=\pi/\omega\kappa$, was found to be independent of the sheet conductivity\cite{Farajollahi2016cmp} (see Appendix \ref{sec:imped}.)
To retain the analogy to conventional waveguides, we treat the interface as a composite object consisting of an ideal transformer and two phase shifters (Fig.~\ref{fig2}b).
The transformer effectively rescales the junction impedance into $Z_j=Z_0\frac{\sigma_0}{\sigma}$ while the complementary phase shifters provide the requisite phase factor $e^{\pm i\theta}$ to the reflection coefficients.
The junction itself can be replaced by 
a piece of waveguide of the effective length $2a - (\theta / q)$, which is
equivalent to a $T$-junction with impedances $Z_1=-iZ_j\tan(qa-\theta/2)$ and $Z_2=iZ_j\csc(2qa-\theta)$ \cite{Montgomery1948pmc}.
The total reflection coefficient from the left interface [coinciding with Eq.~\eqref{eqn:r_t_FP}] can then be found directly from the standard waveguide-theory formula, 
\begin{equation}
	r=({Z_L-Z_0})/({Z_L+Z_0})\,,
\end{equation}
where the total load impedance, calculated by the rules of parallel and series connections, is  $Z_L=Z_1+(Z_2^{-1}+(Z_1+Z_0)^{-1})^{-1}$. 

\subsection{Narrow junctions}
The F-P formula \eqref{eqn:r_t_FP} remains numerically accurate for junctions as narrow as $a\simeq\lambda_0$.
However, as the two leads get closer, they start to couple capacitively. The system resembles a leaky capacitor. 
The capacitive coupling becomes dominant when the junction is narrow. The cavity resonances remain but their amplitude gets strongly modified.
This can be seen in Fig.~\ref{fig3}c,
where the reflectance for a narrow junction $a=0.01\lambda_0$ deviates greatly from the F-P formula but retains the same resonant locations.
In the following we present key intermediate steps in calculating the analytical form of this modified reflectance, leaving the detailed derivation to the Appendix.

We start by discussing the general form of  $r$ for a narrow junction.
The reflection coefficient $r$ can be found by comparing the incident in-plane plasmon field $\mathbf{E_i}=E_0 e^{iq_0x}\hat{x}$ with the asymptotic form of the scattered field $\mathbf{E_s}\simeq -rE_0e^{-iq_0x}\hat{x}$ at large distances from the junction.
The scattered field $E_s$
can be completely determined by the total field $E$ inside the junction  via the Green's function $G$,
\begin{equation}
	E_s\left(|x|>a\right)=\frac{\Delta\sigma}{\sigma_0}\int_{-a}^{a}dx'\,G(x-x')E(x')\,,
	\label{eqn:Es_sheet}
\end{equation}
where  $\Delta\sigma\equiv\sigma-\sigma_0$, and the Green's function $G$ is the 1D Fourier transform of the dielectric function $\epsilon=1-|q|/q_0$ of the 2D sheet,
\begin{equation}
	G(x)=\int_{-\infty}^{\infty}\frac{dq}{2\pi}\,e^{iqx}\epsilon^{-1}(q)\,.
	\label{eqn:Greens}
\end{equation}
By analyzing the asymptotic form of Eq.~\eqref{eqn:Greens}, one arrives at the relation
\begin{equation}
	r=i\frac{\Delta\sigma}{\sigma_0}\frac{q_0}{E_0}\int_{-a}^{a}dx\,E(x)e^{iq_0x}\,,
	\label{eqn:r_Eg}
\end{equation}
which is valid for junctions of \textit{any} width. For example, it can be used to evaluate the reflection coefficient numerically (see Appendix \ref{sec:numerical}.)
For narrow junctions, Eq.~\eqref{eqn:r_Eg} can be simplified and written in terms of the voltage drop across the junction $V=\int_{-a}^{a}dx\,E(x)$,
\begin{equation}
	r\simeq i\frac{\Delta\sigma}{\sigma_0}q_0\frac{V}{E_0}\,,\quad (a\ll\lambda_0)
	\label{eqn:r_V}
\end{equation}
since $e^{iq_0x}\approx 1$ in the junction.

\begin{figure}[t]
	\centering
	\includegraphics[width=0.4\linewidth]{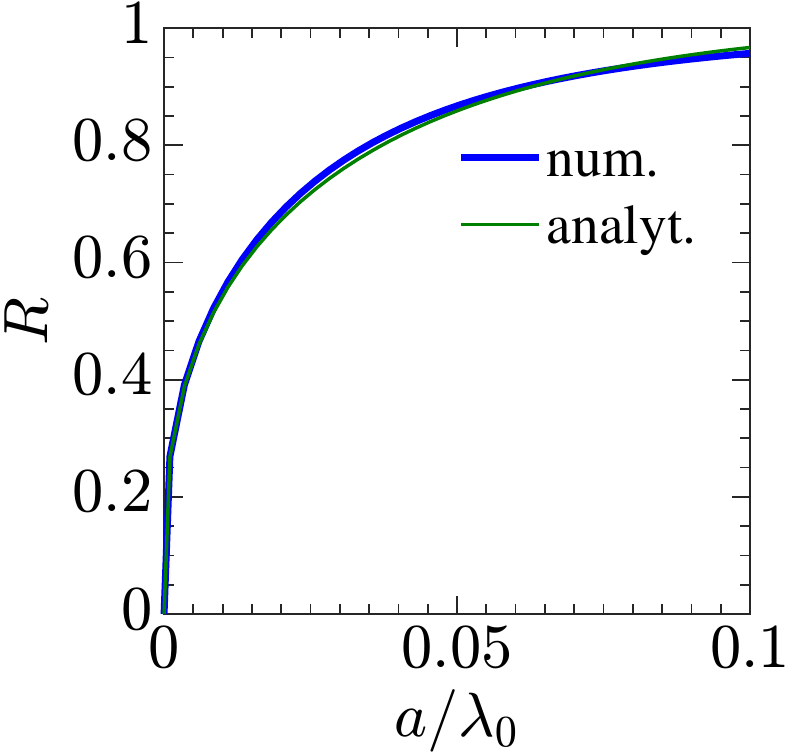}
	\llap{\raisebox{0.085\linewidth}{\includegraphics[height=0.085\linewidth]{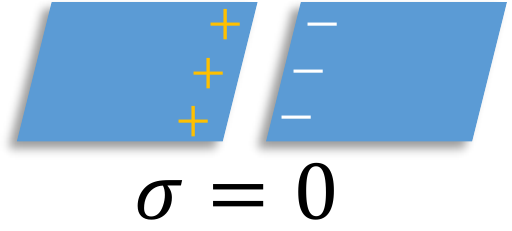}\hspace{0.07\linewidth}}}
	\caption{
		Reflectance of a narrow vacuum gap in a lossless sheet.
	}
	\label{fig4}
\end{figure}

Equation \eqref{eqn:r_V} has simple solutions in limiting cases. 
In the perturbative limit $\sigma\lesssim\sigma_0$, the current density $j(x)$  is approximately constant across the junction, which implies that the field inside the junction is $E\simeq\frac{\sigma_0}{\sigma} E_0$.
This yields the reflection coefficient 
\begin{equation}
	r=2iq_0a\frac{\Delta\sigma}{\sigma}\,,\quad(|\Delta\sigma|\ll\sigma_0)
	\label{eqn:r_perturb}
\end{equation}
in agreement with Ref.\citenum{Fei2013epp}.
The reflection coefficient depends linearly on the strength of the perturbation $\Delta\sigma$, as expected.
In fact, the perturbative expression is also valid for any $\sigma>\sigma_0$ as long as the resulting reflection coefficient is small,  $|r|\ll1$.
However, it is clearly inadequate as $\sigma$ is decreased toward zero, i.e., the case of a vacuum gap, where the perturbative $r$ diverges [Eq.~\eqref{eqn:r_perturb}]. 
For such a vacuum gap, the current $j$ is stopped and completely reflected by the gap, i.e., the two edges of the junction act as a (non-leaky) capacitor.
By considering the charge conservation equation on either edge of the capacitor, $j_i+j_r=\sigma_0 E_0(1-r)=\dot{Q}=-i\omega V C$,
the reflection coefficient for the vacuum gap can be found:
\begin{equation}
	r=\frac{i\kappa}{2\pi C+i\kappa}=\frac{i\pi}{\log\frac{2}{q_0 a}-c+i\pi}\,,\quad (\sigma=0\,,\, a\ll\lambda_0)\,.
	\label{eqn:r_vac}
\end{equation}
Here the junction capacitance per unit length 
\begin{equation}
	C=\frac{\kappa}{2\pi^2}\left(\log\frac{2}{q_0 a}-c\right)\,,\quad c=0.577\ldots
	\label{eqn:C}
\end{equation}
can be derived by considering the charge distribution on perfectly conducting leads (see Appendix \ref{sec:C}.)  
The sublinear dependence on $a$ is due to the 2D geometry of the system, where the fringing fields, i.e., fields exterior to the sheet, dominate.
Equation \eqref{eqn:r_vac} is in good agreement with our numerical result (Fig.~\ref{fig4}) and with Ref.\citenum{Garcia-Pomar2013sgp}.
It can also be simply explained by the equivalent circuit for the system, a resistor-capacitor-resistor series with impedances $Z_0$, $Z_c=i/\omega C$, and $Z_0$.
The circuit yields the load impedance $Z_L = Z_c + Z_0$, so that  $r=Z_c/(Z_c+2Z_0)$, matching Eq.~\eqref{eqn:r_vac}.

For intermediate conductivities $0<\sigma<\sigma_0$, the capacitive component is diminished compared to the vacuum gap, but the junction can now host plasmonic resonances.
This is seen in the modified charge conservation equation,
\begin{equation}
	\sigma_0 E_0(1-r)=-i\omega V C'+
	\int_{-a}^{a}dx\,\frac{\sigma E(x)}{\pi\sqrt{a^2-x^2}}\,,\quad C'= -\frac{\Delta\sigma}{\sigma_0}C\,,
	\label{eqn:Shockley-Ramo}
\end{equation}
where the capacitance acquires a correction factor $-\Delta\sigma/\sigma_0$ which is $1$ for the vacuum gap but decreases to 0 for a uniform sheet where $\sigma=\sigma_0$.
The additional integral
represents the Shockley-Ramo image current in the leads induced by the current in the junction \cite{Ryzhii2003asd}.
The field in the junction can be expanded in the basis of resonant eigenmodes
\begin{equation}
	E\left(|x|<a\right)=\frac{V}{2a}\left(1+\sum_{n=1}^{\infty} {b_n} f_n(x)\right)\,.
	\label{eqn:E_gap}
\end{equation}
The coefficients
\begin{equation}
	b_n=\frac{2a}{1+\frac{\sigma}{\Delta\sigma}\frac{q_n}{q_0}}\frac{\int_{-a}^a dx\,\frac{f_n(x)}{\pi\sqrt{a^2-x^2}}}{\int_{-a}^a dx\,f_n^2(x)}
	\label{eqn:b_n}
\end{equation}
diverge 
whenever the plasmon wavevector $q$ inside the junction matches the resonant wavevector $q_n$, 
and the field $E$ becomes dominated by the $n$-th resonant mode $f_n$.
The wavevector $q_n$ of the resonant field is again determined by the resonance condition $2q_n a=n\pi+\theta$ as in the case of wide junctions. 
For narrow junctions, all resonances occur at $\sigma_0/\sigma\gg1$, so that $\theta\approx\pi/4$ and $q_n=\frac{\pi}{2a}(n+\frac14)$.
The exact forms of the dimensionless resonant fields $f_n$ has to be calculated numerically (the simple forms described in Ref. \citenum{Ryzhii2003asd} are inaccurate, see Appendix \ref{sec:prev_work}), but they quickly approach $\cos(q_n x-\frac{1 }{2}n\pi)$ as $n$ is increased \cite{Ryzhii2003asd,Ryzhii2004pos}.
The first three eigenmodes $f_n(x)$, $n=1,2,3$, are plotted in Fig.~\ref{fig1} and also in Fig.~\ref{fig:eigen}.

\begin{figure*}[t]
	\centering
	\includegraphics[width=.85\linewidth]{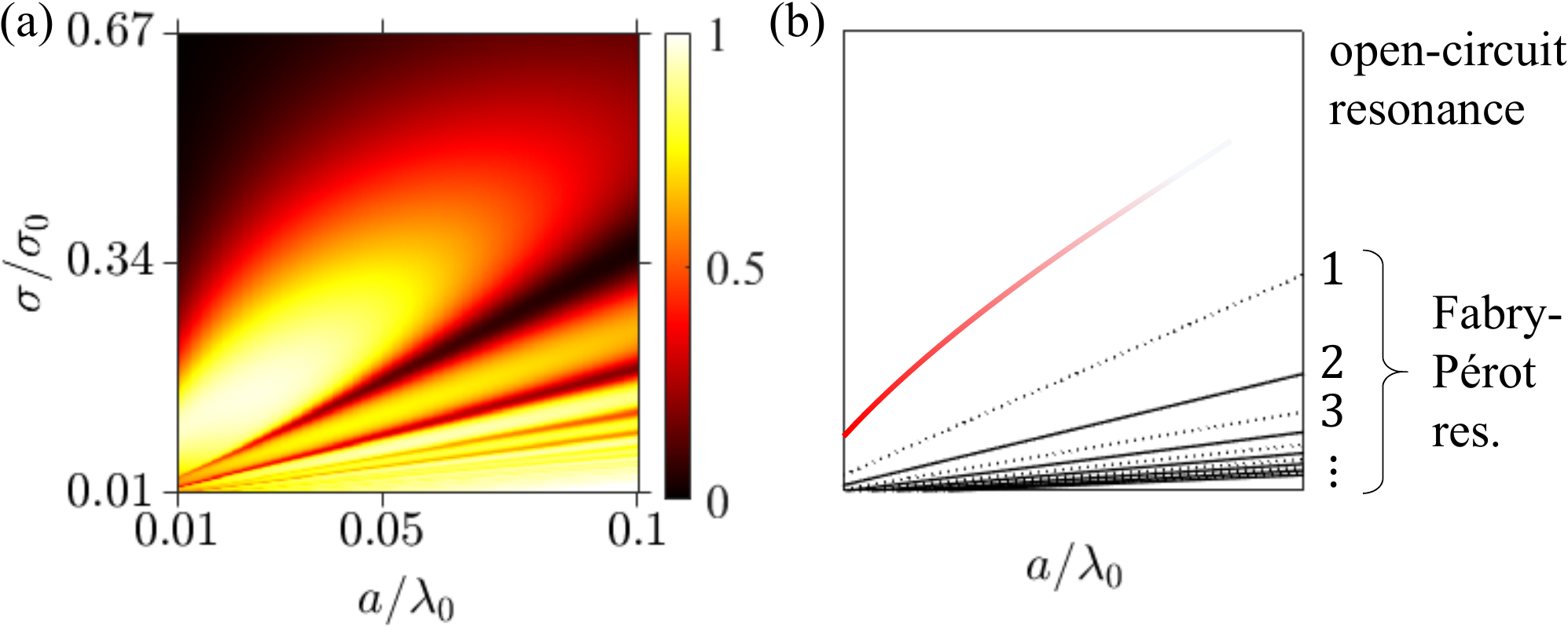}
	\caption{
		(a) False color plot of the reflectance of the junction for $\gamma=0.05$.
		(b) Schematic diagram of the location of the open-circuit and the F-P resonances.
		The red curve is predicted by Eq.~\eqref{eqn:open_circ}.
	}
	\label{fig5}
\end{figure*}

The reflection coefficient can be found by substituting the field Eq.~\eqref{eqn:E_gap} into Eq.~\eqref{eqn:Shockley-Ramo}, then using the general form of $r$  [Eq.~\eqref{eqn:r_V}] to obtain
\begin{eqnarray}
r^{-1}=1-i\frac{2\pi}{\kappa}C-i\frac{1}{2q_0 a}\frac{\sigma}{\Delta\sigma}\left(1+\sum_{n=1}^{\infty}b_n \int_{-a}^a dx\,\frac{f_n(x)}{\pi\sqrt{a^2-x^2}}\right)\,.
\label{eqn:r_analyt}
\end{eqnarray}
At a resonance, $r^{-1}$ diverges for zero damping and has a maximum for finite damping, so the reflectance has a local minimum, similar to the case of wide junctions.
The main difference from a wide junction comes from the capacitive coupling, which gives rise to the non-resonant terms in Eq.~\eqref{eqn:r_analyt}. 
These terms vary smoothly with $\sigma$ and $a$,
giving the resonances a Fano shape and causing $r$ to deviate significantly from the F-P formula, as shown in Fig.~\ref{fig3}c and d.
Another distinction from wide junctions is that even and odd resonances become very different in strength.
Numerical results show that the odd resonances are more narrow and less strong than the even ones.
This can be seen more clearly in the analytical formula for $r$ calculated under the approximation 
\begin{equation}
	f_n=\cos(q_n x-\frac{1 }{2}n\pi)\,,\quad q_n=\frac{\pi}{2a}\left(n+\frac14\right)\,.
\end{equation}
In this case the odd modes disappear completely due to the integral in Eqs.~\eqref{eqn:b_n} and \eqref{eqn:r_analyt} yielding $b_n=0$ for an odd $f_n(x)$.
The residual presence of the odd modes in the numerical results is due to the incident field $E_i=E_0 e^{iq_0x}$ containing both even and odd components.
Except for this difference,  the analytical approximation  agrees well with the numerical result, 
and the agreement gets better at larger $n$ where the approximation for $f_n$ becomes more accurate.

Equivalent circuits are again helpful in understanding Eq.~\eqref{eqn:r_analyt}.
Consider the impedance of a $2a$-long piece of plasmonic waveguide $Z=-4iZ_0\tan qa$,\cite{Montgomery1948pmc} which can be expanded into the resonant form  $Z^{-1}=(-4iZ_0qa)^{-1}+i\frac{qa}{2Z_0}\sum_{n}(q^2a^2-\pi^2n^2)^{-1}$.
This inverse impedance has the same structure as 
a network of parallel $LC$ circuits (Fig.~\ref{fig2}c), which is in fact how resonant cavities are conventionally described.
The impedance of the $LC$ network is $Z^{-1}=(-i\omega{L_0})^{-1}+i\omega\sum_n (\omega^2 L_n-C_n^{-1})^{-1}$, which yields
${L_0}=2i a/\sigma\omega$, $L_n=\frac12 L_0$, and $C_n=qa\kappa/2\pi^3n^2$.
The load impedance of the junction, including the capacitor, the $LC$ network and the right half-plane is then $Z_L=ZZ_c/(Z+Z_c)+Z_0$, which yields the reflection coefficient
\begin{equation}
	r^{-1}=1-i\frac{2\pi}{\kappa} C+\frac{i}{2qa}+iqa\sum_{n=1}^{\infty}\frac{1}{q^2a^2-\pi^2n^2}\,.
\end{equation}
The first three terms represent the low $q$ or dc response and match those in Eq.~\eqref{eqn:r_analyt} (under the limit $\sigma\ll\sigma_0$).
The strength of the resonances does not match Eq.~\eqref{eqn:r_analyt} but this is expected,
as the effect of capacitive coupling on the resonant modes was not accounted for.
To correctly describe the junction $L_n$ and $C_n$ will have to be modified, which is effectively accomplished by Eq.~\eqref{eqn:b_n}.
The resonant condition $qa=\pi n$ is however qualitatively correct (apart from the phase shift $\theta$) and predicts the disappearance of the odd modes.



As in the case of wide junctions, the high-order resonances quickly diminish in the presence of plasmonic damping (Fig.~\ref{fig3}d), so these resonances are not robust enough for plasmonics applications.
However, there is a prominent peak in the reflectance that occurs apart from the resonances and is persistent under damping,
making it a promising candidate for controlling a compact plasmonic switch. 
The physical origin of this peak is as follows. If the field in the junction is approximately uniform, 
the junction reduces to a parallel $LC$ circuit (Fig.~\ref{fig2}d) with impedances $Z_C=i/\omega C$ and $Z_{L_0}=-i\omega L_0=\frac{2a}{\sigma}$, the later resembling the usual dc formula for resistance. 
When the inverse impedances, i.e., admittances $Z_C^{-1}$ and $Z_{L_0}^{-1}$ cancel, the total impedance of the junction diverges and the reflectance is unity. We call this situation the open-circuit resonance.
It occurs when
\begin{equation}
	\frac{\sigma}{\sigma_0-\sigma}\frac{\lambda_0}{a}=4\left(\log\frac{2}{q_0 a}-c\right)\,.
	\label{eqn:open_circ}
\end{equation}
The open-circuit resonance arises only for narrow junctions because it requires capacitive coupling of the leads.
As the junction width increases, the capacitive coupling gradually weakens, and the junction evolves towards a regular F-P resonator, as shown in Fig.~\ref{fig5}.



\subsection{Power absorption}
Low-conductivity junctions based on narrow slots in metallic films  or 2DEGs under split-gates have also been proposed as basic units of terahertz detectors.
They function by converting photons into plasmons, which are then 
detected as a dc current either through  nonlinear optical effects\cite{Dyakonov1996dmf,Knap2009fet} or  thermoelectric effects \cite{Tong2015aeg,Lundeberg2017tdi,Woessner2017edh}.
The thermocurrent generated by the plasmons is proportional to the total power $P$ absorbed by the system.
Here we show that $P$ 
is related to the plasmon reflectance $R$ of  the junction, so that the analytical formula Eq.~\eqref{eqn:r_analyt} can be used to predict the detection efficiency.
The photon to plasmon conversion can be understood from our formulas above since the external terahertz field effectively replaces the field of the incident plasmon wave.
For a narrow gap both fields are approximately constant over the junction, so that they launch plasmons from the junction with the same efficiency.



\begin{figure*}[t]
	\centering
	\includegraphics[width=.75\linewidth]{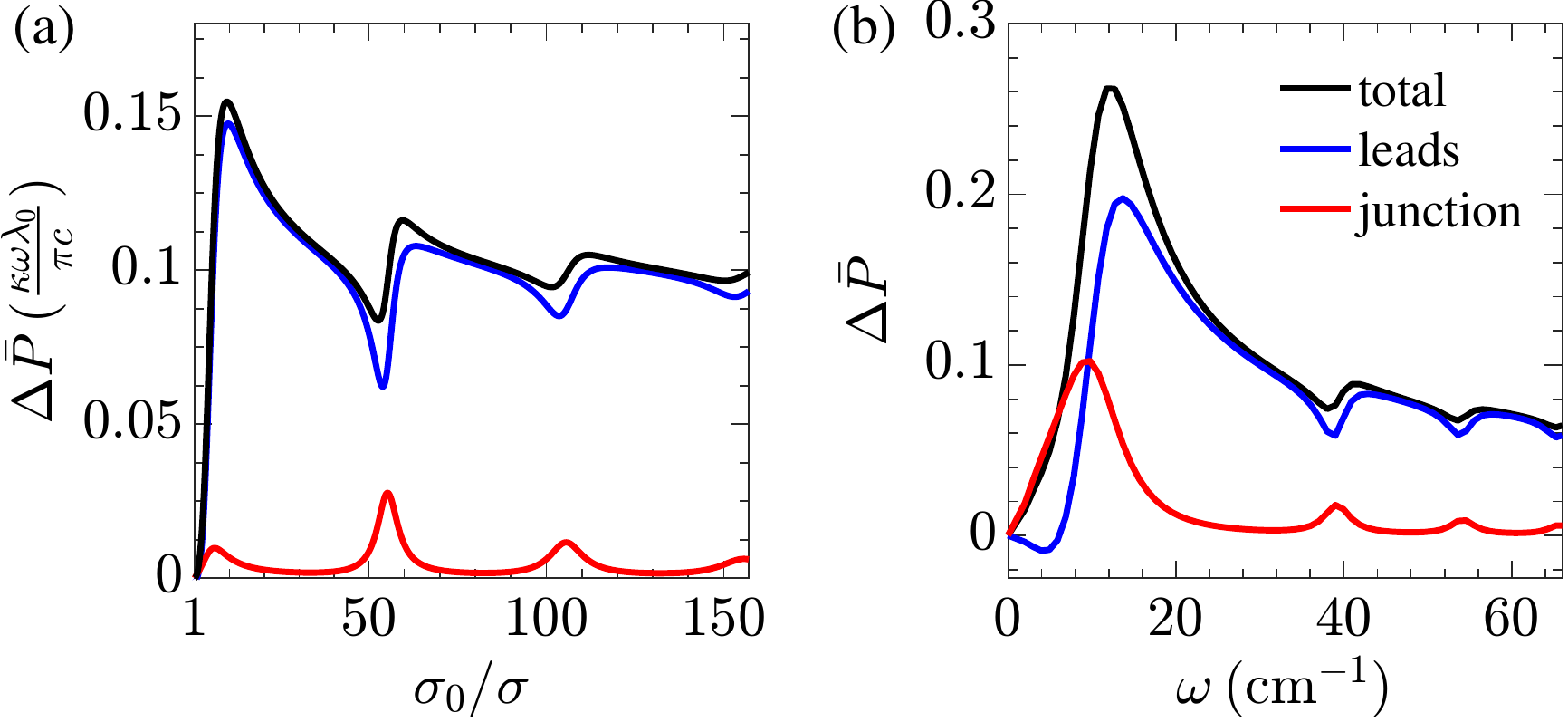}
	\caption{
		(a) Excess power absoption $\Delta\bar{P}$ for $a=0.01\lambda_0$ and $\gamma=0.05$.	
		The power absorbed by the leads (blue) is proportional to the reflectance $R$.
		The power absorbed by the junction (red) has a maximum  at the open-circuit and the F-P resonances.
		(b) Excess power absoption spectrum for a junction of width $2a=1000\,\mathrm{nm}$ in graphene on hBN. 
		The background conductivity is assumed to be Drude like, $\sigma_0=\frac{iD_0}{\pi(\omega+i\nu)}$ with $\nu=3\,\mathrm{cm^{-1}}$.
		The Drude weight $D_0=\frac{e^2}{\pi\hbar^2}\mu$ is calculated at chemical potential $\mu=0.12 \mathrm{eV}$, 
		the conductivity ratio is $D/D_0=0.03$.
		At low frequencies $\omega\sim\nu$ the conductivity is dominated by damping, resulting in a negative $\Delta\bar{P}_l$, i.e., the absorption  in the leads is smaller than the background value. 
	}
	\label{fig6}
\end{figure*}

To quantify the extra absorbed power when the junction is present compared to a uniform 2DEG sheet, we define the dimensionless excess power absorption,
\begin{equation}
	\Delta\bar{P}=\frac{\int_{-\infty}^{\infty}dx\,\left(P(x)-P_0\right)}{\lambda_0\cdot P_\mathrm{inc}}
\end{equation}
where
\begin{equation}
	P(x)=\mathrm{Re}\,\sigma(x)|E(x)|^2
\end{equation}
is the local power absorption per unit area,  $P_0=\mathrm{Re}\,\sigma_0|E_0|^2$ is the same quantity for a uniform system without a junction, and  $P_\mathrm{inc}=\frac{c}{4\pi}|E_0|^2$ is the incident power per unit area.
We further divide $\Delta\bar{P}$ into two parts, absorption by the junction $(\Delta\bar{P}_j)$ and by the leads $(\Delta\bar{P}_l)$. 
The former is determined by the junction conductivity as well as the strength of the resonant fields, while the latter is determined by the strength of plasmon emission.
The power absorption of the junction has a peak at each resonance where the resonant field is strong.
It is also peaked at the open-circuit resonance, where the field in the junction is approximately constant and proportional to $r$, $E\simeq V/2a\propto r$, so that the peak in $R$ is mirrored in $\Delta\bar{P}_j$.
The power absorbed by the leads is determined by the scattered plasmon field which quickly approach $-re^{iq_0|x|}$ away from the junction.
Hence $\Delta\bar{P}_l$ is approximately proportional to the reflectance.
As shown in Fig.~\ref{fig6}a, $\Delta\bar{P}_l$  is very similar to the reflectance shown in Fig.~\ref{fig3}d.
The resonances occur at the same locations but the odd modes disappear, due to the incident field being even in $x$.

A similar resonant structure is found when the frequency $\omega$ instead of the junction conductivity $\sigma$ is varied, as shown in Fig.~\ref{fig6}b.
Varying $\omega$ changes  the background wavelength $\lambda_0$ and thus the ratio $a/\lambda_0$.
The enhanced $\Delta\bar{P}_j$ at low frequencies is due to our assumption of a Drude-like conductivity,
$\sigma={iD}/{\pi(\omega+i\nu)}$
where the damping rate $\nu$ is taken to be constant, so that $\mathrm{Re}\,\sigma\propto\frac{\nu}{\omega^2+\nu^2}$ has a maximum at $\omega=0$.
The Drude weight $D$ is assumed to be proportional to the carrier density and independent of frequency.
At low frequencies using the approximations $E\simeq V/2a$ and  $E_s\simeq -re^{iq_0|x|}$, we obtain
\begin{equation}
	\Delta\bar{P}_j\simeq\frac{2}{\pi \kappa g^2}\frac{D_0^2}{c a}\frac{\nu}{(\omega^2+\nu^2)^2}R\,,\quad
	\Delta\bar{P}_l\simeq\frac{4}{\pi}\frac{D_0}{c}\frac{\omega}{(\omega^2+\nu^2)}R\,,
\end{equation}
where $c$ is the speed of light and $g\equiv D/D_0-1$.
Using these expressions, the optimal operating frequency can be found given the device parameters.

\section{Discussion}
As previously mentioned, in the THz range where the plasmon wavelength $\lambda_0$ is often larger than the distance $d$ between the 2D sheet and the gate, the effects of screening cannot be neglected in general.
However, if the junction is narrow enough such that $a\ll d$, the capacitive coupling that gave rise to the absorption peak still occur.
Screening effectively replaces the long range cutoff $1/q_0$ by $d$ in the capacitance, Eq.~\eqref{eqn:C}, causing a small shift in the peak absorption frequency.


Besides electrostatically gated graphene or
2DEGs, tunable conductivity profile needed for the plasmon-reflecting 1D junction can be also  realized with layered high-$T_c$ superconductors by varying the temperature.
Plasmon reflection in such a system would also yield information about local optical conductivity, which can be difficult to measure by other means.
If the conductivity contrast is large, which is the case of our primary interest, the system becomes a weak link, i.e., a Josephson junction. Such a junction can be fabricated
by lowering the $T_c$ locally using focused ion beams or by etching and reducing the number of layers \cite{Cybart2015njs}.
An estimate of the required parameters are as follows.
Assume a BCS-like superconductor film has a thickness $l=10\,\mathrm{nm}$, interlayer Josephson plasmon frequency $\omega_c=50\,\mathrm{cm^{-1}}$, lattice dielectric constant $\epsilon_\infty=27$, anisotropy $\gamma=17.5$, and is placed on a strontium titanate (STO) substrate.
The low-frequency plasmon dispersion is then $\omega\simeq\gamma\omega_c\sqrt{(1+2\kappa/\epsilon_\infty ql)^{-1}}$ \cite{Stinson2014ini}.
At a frequency $\omega=0.25\,\mathrm{THz}$, the effective dielectric constant of STO is $\kappa\simeq1000$,
and so the plasmon wavelength $\lambda_0\simeq10\,\mu \mathrm{ m}$.
If  the conductivity  inside the central region is $\sigma=0.1\sigma_0$, then the strongest reflection/optimal $\mathrm{THz}$ detection occurs for a junction of width $2a\sim 200\,\mathrm{nm}$, which is quite practical.

\section{Conclusion}
We presented a theory of plasmonic interaction with a conductivity dip and showed that a narrow junction is a minimalistic yet robust plasmonic switch.
Our analytical results and the associated physical insights into the plasmonic interaction with a 1D defect may be useful for the design of nanoscale plasmonic reflectors as well as terahertz detectors.

\titleformat{\section}{\bfseries}{\appendixname~\thesection.}{0.5em}{}

\appendix
\section{Fabry-P\'{e}rot formula for wide junctions}

At the interface $x=0$ of two conductive sheets with conductivities $\sigma_1$ and $\sigma_2$, 
a normally incident plasmon plane wave from sheet $1$ has the following reflection and transmission coefficients\cite{Rejaei2015ssp}
\begin{equation}
r_{12}=e^{i\theta_{12}} \frac{\sigma_1-\sigma_2}{\sigma_1+\sigma_2}\,,\quad
t_{12}=\frac{2\sqrt{\sigma_1\sigma_2}}{\sigma_1+\sigma_2}\,,\quad
\theta_{12}=\frac{\pi}{4}-\frac2\pi\int_0^\infty du\,\frac{\tan^{-1}( \frac{\sigma_2}{\sigma_1}u)}{u^2+1}\,.
\label{eqn:r_interface}
\end{equation}
They have the properties $t_{12}=t_{21}$ and $\theta_{12}=-\theta_{21}$.
Our junction has two interfaces between conductivities $\sigma_0$ and $\sigma$.
Assigning $\sigma_1=\sigma_0$ and $\sigma_2=\sigma$,
the total reflection and transmission from the left edge of the junction can be found by considering the multiple reflections within the junction,
\begin{equation}
r=r_{12}+\left(t_{12}r_{21}e^{i2\phi}t_{21}+t_{12}r_{21}^3e^{i4\phi}t_{21}+...\right)=r_{12}+\frac{t_{12}r_{21}e^{i2\phi}t_{21}}{1-r_{21}^2e^{i2\phi}}\,,
\label{eqn:r_multi_reflect}
\end{equation}
\begin{equation}
t=t_{12}e^{i\phi}t_{21}+t_{12}r_{21}^2e^{i3\phi}t_{21}+...=\frac{t_{12}e^{i\phi}t_{21}}{1-r_{21}^2e^{i2\phi}}\,,
\label{eqn:t_multi_reflect}
\end{equation}
where $\phi=2qa$ is the phase the plasmon wave accumulates for crossing the junction once.
The F-P formula 
\begin{equation}
r_{\mathrm{FP}}^{-1}=\frac{\sigma_0^2+\sigma^2}{\sigma_0^2-\sigma^2}+i\frac{2\sigma_0\sigma}{\sigma_0^2-\sigma^2}\cot(\phi+\theta)
\label{eqn:r_FP}
\end{equation}
can be obtained by substituting Eq.~\eqref{eqn:r_interface} into Eq.~\eqref{eqn:r_multi_reflect}.
Note that the formula above was derived assuming the left interface to be the origin $x=0$.
If the origin is set at the center of the junction instead, which sits at a distance $a$ from the left interface, the reflection and transmission coefficients Eqs.~\eqref{eqn:r_multi_reflect}, \eqref{eqn:t_multi_reflect} and \eqref{eqn:r_FP} will acquire an additional phase factor $e^{-i\phi}$.

\section{Definition of impedances in the equivalent circuits}
\label{sec:imped}

\begin{figure}[t]
	\centering
	\includegraphics[width=0.7\linewidth]{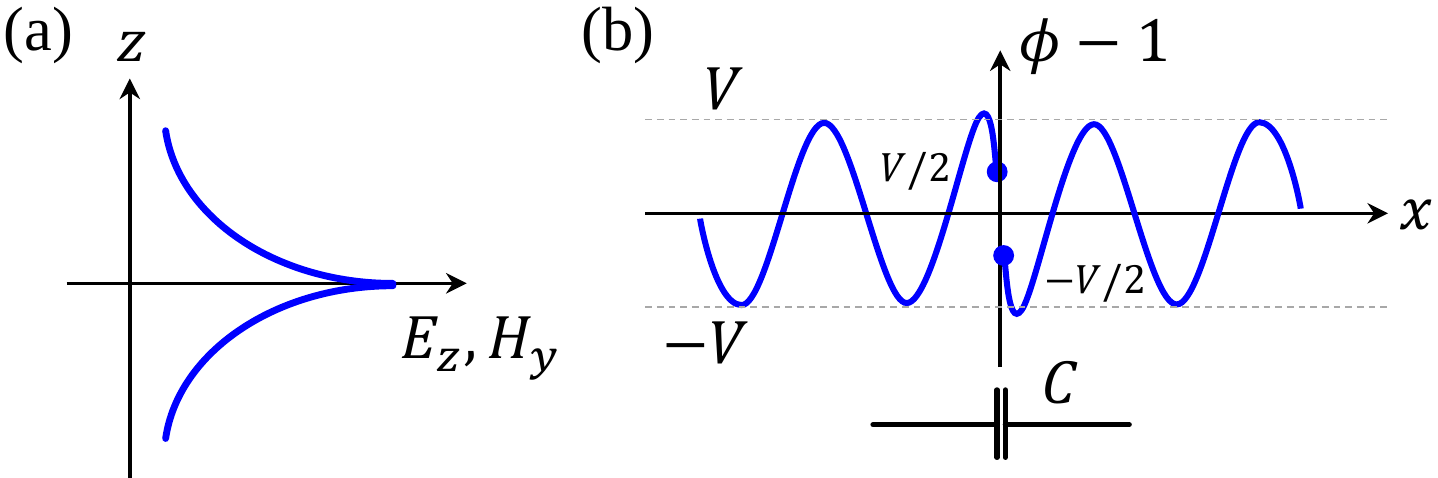}
	\caption{
		(a) Distribution of $E_z$ and $H_y$ outside the sheet.
		(b) Schematic diagram of the potential distribution for the case of a narrow vacuum gap.
		The large distance oscillations have amplitude $V$, while the potential right next to the gap oscillates with amplitude $V/2$.
	}
	\label{fig:imped}
\end{figure}

There is always some arbitrariness~\cite{Montgomery1948pmc} in the value of the characteristic impedance $Z_0 = v / i$ of a waveguide, which derives from the freedom of choosing the definition of ``voltage'' $v$ and ``current'' $i$ of a propagating wave.
The only constraint is that the product of the two is
proportional to the power $P$ transmitted by such a wave, more precisely, that
\begin{equation}
P=\frac12\, \mathrm{Re}\left(v^\ast i\right)\,.
\end{equation}
In our case the propagating wave is a plasmon. The plasmon travelling in the $x$-direction can be described by the field amplitudes
\begin{equation}
{E_x}=e^{iq_0x}e^{-q_0|z|}\,,\quad E_z=iE_x\,,\quad H_y=\frac{2\pi}{c}\sigma_0E_x\,,
\end{equation}
which decay exponentially away from the plane, see Fig.~\ref{fig:imped}(a).
The power transmitted by the plasmon (per unit length in $y$) is given by the integrated Poynting vector,
\begin{equation}
P=\frac12\, \mathrm{Re}\int_{-\infty}^\infty dz\,\frac{c}{4\pi}H_y^\ast E_z=\frac14\, \mathrm{Re}(j^\ast \phi)\,.
\end{equation}
Here  $\phi$ is the electric potential on the plane,
\begin{equation}
\phi=\int_{0}^{\infty}dz\,E_z\,,
\end{equation}
and $j=\sigma_0 E_x$ is the current density.
In this work we define our effective current to be equal to the physical current, $i=j$.
This choice leads to a simple form of the current conservation in our equivalent circuits
and fixes the effective voltage $v$ to be one-half of the physical potential:
\begin{equation}
v=\frac12\phi\,,\quad i=j\,.
\label{eqn:v}
\end{equation}
The characteristic impedance of the sheet is then
\begin{equation}
Z_0=\frac12\frac{\phi}{j}=\frac{\pi}{\omega\kappa}\,.
\label{eqn:Z_0}
\end{equation}
One further application of this result is to a wide but finite strip.
If this strip has the unperturbed sheet conductivity $\sigma = \sigma_0$ and the total length $2a \gg q_0^{-1}$ in the $x$-direction,
then it is equivalent to a T-junction shown in Fig.~1(b) of the main text.
The horizontal legs have the impedance~\cite{Montgomery1948pmc}
\begin{equation}
Z_1 = -i Z_0\tan q_0 a
\label{eqn:Z_1}
\end{equation}
each and the vertical leg has the impedance
\begin{equation}
Z_2 = i Z_0\csc q_0 a\,.
\label{eqn:Z_2}
\end{equation}

However, using Eqs.~\eqref{eqn:Z_1} and \eqref{eqn:Z_2} for a narrow strip $2a \ll q_0^{-1}$ would be wrong
because in that case the electric and magnetic fields have a much more complicated distribution than what is sketched in Fig.~\ref{fig:imped}(a). 
Nevertheless, a simple result is obtained for a narrow vacuum gap, $\sigma = 0$, whose interaction with the plasmons is governed by
the capacitive coupling of the two leads.
As shown in the main text [see also Eq.~\eqref{eqn:r_vac} below], the correct plasmon reflection coefficient is obtained if we assume
that the impedance of this type of junction is given by the usual circuit formula for the capacitor,
\begin{equation}
Z_C = \frac1{-i\omega C}\,.
\label{eqn:Z_C}
\end{equation}
Notably, it has no additional factor of one-half, which one would naively guess based on Eq.~\eqref{eqn:v}.
The reason for its absence is the difference between the ``near-field'' potential $V$ and the ``far-field'' potential $\phi$.
Potential $\phi$ in our definiton of the impedance refers to the \textit{asymptotic} amplitude of the plasmon waves.
Our formula for the the reflection coefficient [Eq.~\eqref{eqn:r_narrow} below]
indicates that for the incident wave $\phi_i=e^{iq_0 x}$, 
the scattered potential at large distances oscillates with an amplitude $V$.
On the other hand, since $V$ is also the potential difference across the gap,
its amplitude just outside the gap oscillates with an amplitude $V/2$, as shown in Fig.~\ref{fig:imped}(b).
This factor-of-two difference between the far-field and near-field amplitudes cancels the factor of one-half in
the definition of the effective voltage $v = \phi / 2$, leading to the standard form of Eq.~\eqref{eqn:Z_C}.

The same cancellation occurs also if the short strip has a nonzero sheet conductivity $\sigma$.
In other words, the correct result for the reflection coefficient is obtained if impedances $Z_1$ and $Z_2$ of the
T-junction are multiplied by the factor of two.
As far as the vertical leg is concerned, this modification has virtually no effect because it already has a large impedance $Z_2 \gg Z_1$,
and so can be cut from the circuit.
But the two horizontal legs now combine to the total impedance of 
$Z \simeq 4 Z_1 = -4 i Z_0\tan q a$, as mentioned in the main text.
The last expression can be further approximated by
\begin{equation}
Z \simeq -4 i Z_0 q a = -4i\, \frac{\pi}{\omega\kappa}\, \frac{i \omega \kappa}{2\pi \sigma}\, a
= \frac{2a}{\sigma}\,,
\qquad q a \ll 1\,,
\label{eqn:Z_2a}
\end{equation}
which is the usual formula for dc resistance of a conductor, similar to Eq.~\eqref{eqn:Z_C} being the standard formula for a capacitor.

\section{Reflection coefficient of general 1D inhomogeneities}

\textbf{Plasmon equation.} To make the paper self-contained, we begin this section with a derivation of the
plasmon reflection coefficient for a 1D inhomogeneity.
The quantity of interest is the electric potential $\phi$ on the sheet.
Within the quasistatic approximation, which is valid when all distances involved are much smaller than the radian length $c/\omega$, the potential is given by the Coulomb law,
\begin{equation}
\phi(\textbf{r}) = \phi_\mathrm{ext}(\textbf{r})
+ (V \ast \delta\rho)(\textbf{r})\,,\quad \textbf{r}=(x,y)\,.
\label{eqn:coulomb_law}
\end{equation}
Here and below the time dependence enters via the factor $e^{-i\omega t}$, which is implicit.
An external potential $\phi_\mathrm{ext}$ induces a charge distribution $\delta \rho$ on the sheet, which in turn creates a potential $V\ast\delta\rho$.
Here $V=1/\kappa {r}$ is the Coulomb kernel,
$\kappa$ is the  dielectric constant of the environment exterior to the sheet, 
and $\ast$ denotes convolution,
\begin{equation}
(A \ast B)(\textbf{r}) \equiv \int d^2 r'\, A(\textbf{r}-\textbf{r}')B(\textbf{r}')\,.
\end{equation}
Using the continuity equation $\partial_t\delta\rho+\nabla\cdot\mathbf{j}=0$, which is equivalent to
\begin{equation}
-i\omega\delta\rho+\nabla\cdot(\sigma\mathbf{E})=0\,,
\end{equation}
where $\sigma = \sigma(\textbf{r})$ is the local sheet conductivity,
we recast Eq.~\eqref{eqn:coulomb_law} into
\begin{equation}
\phi(\textbf{r}) = \phi_\mathrm{ext}(\textbf{r})
- V(\textbf{r})\ast{\nabla}\cdot
\left(\frac{\sigma(\textbf{r})}{i\omega}{\nabla}\phi(\textbf{r})\right)\, .
\label{eqn:plasmon_eq}
\end{equation}
This is the principal equation governing the propagation of plasmons in a sheet.
In general, it has to be solved numerically.

For the case of a uniform sheet with conductivity $\sigma(\mathbf{r})=\sigma_0$,
Eq.~\eqref{eqn:plasmon_eq} can be solved analytically in the momentum space,
\begin{equation}
\tilde{\phi}(\mathbf{k})=\tilde{\phi}_\mathrm{ext}(\mathbf{k})+\tilde{V}(\mathbf{k})\cdot q^2\frac{\sigma}{i\omega}\tilde{\phi}(\mathbf{k})\,,
\label{eqn:plasmon_eq_q_uniform}
\end{equation}
where $\tilde{V}(\mathbf{k})={2\pi}/{\kappa|\mathbf{k}|}$
and the Fourier transform is defined as
\begin{equation}
\tilde{f}(\mathbf{k})=\int d\mathbf{r}\,f(\mathbf{r}) e^{-i\mathbf{k}\cdot\mathbf{r}}\,,\quad
{f}(\mathbf{r})=\int \frac{d\mathbf{k}}{(2\pi)^2}\,\tilde{f}(\mathbf{k}) e^{i\mathbf{k}\cdot\mathbf{r}}\,.
\end{equation}
Eq.~\eqref{eqn:plasmon_eq_q_uniform} has the form 
\begin{equation}
\tilde{\phi}(\mathbf{k})=\frac{\tilde{\phi}_\mathrm{ext}(\mathbf{k})}{\epsilon(\mathbf{k})}\,,
\label{eqn:phi_epsilon}
\end{equation}
where the dielectric function is define by
\begin{equation}
\epsilon(\mathbf{k})=1-\frac{|\mathbf{k}|}{q_0}\,.
\end{equation}
The plasmon momentum $q_0$ is found by setting $\epsilon$ to zero:
\begin{equation}
q_0=\frac{i\kappa\omega}{2\pi\sigma_0}\,.
\label{eqn:plasmon_q_sigma}
\end{equation}
This quantity is complex when a dissipation is present in the system, e.g., when the dielectric constant of the environment $\kappa$ has a nonzero imaginary part, or when the conductivity $\sigma_0$ has a nonzero real part.
The imaginary part $\mathrm{Im}\, q_0 > 0$ has the physical meaning of the inverse propagation length.
For convenience, we parametrize $q_0$ as
\begin{equation}
q_0=\frac{2\pi}{\lambda_0}(1+i\gamma)\,,
\end{equation}
where $\lambda_0=2\pi/\mathrm{Re}\,q_0$ is the plasmon wavelength and $\gamma=\mathrm{Im}\, q_0/\mathrm{Re}\, q_0$ is the dimensionless damping.
In the absence of the external potential,
one can find (unbounded) solutions $\phi = e^{i q_x x + i q_y y}$
with real $q_y$ and complex $q_x = \sqrt{q_0^2 - q_y^2}$,\, $\mathrm{Im}\, q_x > 0$,
which can be thought of as decaying plane waves that are incident from the far left at some oblique angle.\newline

\noindent\textbf{1D inhomogeneities.} In our problem the plasmon wave $\phi_i = e^{i q_x x + i q_y y}$ impinges upon a 1D inhomogeneity in the  conductivity of the sheet  localized around the $y$-axis.
We parametrize the inhomogeneity as
\begin{equation}
\sigma(x)=\sigma_0\left[1+g(x)\right]
\end{equation}
or
\begin{equation}
g(x)=\frac{\Delta\sigma(x)}{\sigma_0}\,,\quad\Delta\sigma(x)\equiv\sigma(x)-\sigma_0\,.
\end{equation}
As the conductivity and the plasmon wavevector are inversely proportional to each other [Eq.~\eqref{eqn:plasmon_q_sigma}], the inhomogeneity can also be parametrized in terms of the plasmon momentum,
\begin{equation}
\frac{1}{q(x)} = \frac{1+g(x)}{q_0}\,,
\quad q_0 \equiv q(\infty)\,.
\label{eqn:g_def2}
\end{equation}
A spatial variation in the momentum scatters an incident plasmon wave $\phi_i$, 
so that the solution $\phi$ to the plasmon equation Eq.~\eqref{eqn:plasmon_eq} contains
both the incident $\phi_i$ and the scattered (reflected plus transmitted) waves $\psi$.
Setting ${\phi}_\mathrm{ext}(\mathbf{r}) \to 0$ and
$\phi(\mathbf{r}) \to \phi(x) e^{i q_y y}$ in Eq.~\eqref{eqn:plasmon_eq}, where the $e^{iq_yy}$ factor is retained by virtue of translational invariance in the $y$-direction,
we obtain the 1D version of the plasmon equation,
\begin{equation}
\phi(x) = V_1 \ast \left(\dfrac{1 + g(x)}{q_0} q_y^2 \phi(x)
- \partial_x \dfrac{1+g(x)}{q_0} \partial_x\phi(x)
\right)\,.
\label{eqn:plasmon_eq_1d}
\end{equation}
Here $V_1(x)$ is the 1D Coulomb kernel divided by $2\pi$,
\begin{equation}
V_1(x)=\frac1{2\pi}\int_{-\infty}^\infty dy\frac{e^{i q_y y}}{\sqrt{x^2+y^2}} = \frac{K_0(|q_y x|)}{\pi}\,,
\end{equation}
where $K_0(z)$ is the modified Bessel function of the second kind.
Note that $\tilde{V}_1(\mathbf{k}) = 1/|\mathbf{k}|$.

To find the expression for the scattered wave $\psi$ we again go to the momentum space,
where Eq.~\eqref{eqn:plasmon_eq_1d} becomes
\begin{equation}
\left(\tilde{\phi}_i+\tilde{\psi}\right)\epsilon=\frac{1}{q_0}\tilde{V}_1\left(q_y^2(\widetilde{g\phi})-
(\widetilde{\partial_xg\partial_x\phi})\right)\,.
\end{equation}
The incident field $\phi_i$ is a solution of the homogeneous equation Eq.~\eqref{eqn:phi_epsilon}, so $\tilde{\phi}_i\epsilon=0$,
while the inverse dielectric function $\epsilon^{-1}$ has the meaning of the Fourier transformed Green's function, $\tilde{G}=\epsilon^{-1}$.
The equation for the scattered wave in real space is then
\begin{equation}
\psi(x) = \frac1{q_0}(G\ast V_1) \ast \left[q_y^2 g(x)  \phi(x)
- \partial_x g(x) \partial_x \phi(x)\right].
\label{eqn:plasmon_eq_1d_psi}
\end{equation}
where
the Green's function is
\begin{equation}
G(x, q_y) = \int_{-\infty}^{\infty} \frac{dk}{2\pi} e^{i k x}
\epsilon^{-1}\left(\sqrt{k^2 + q_y^2}\right)\,.
\end{equation} 
The integrand of the Green's function has two poles at $k=\pm q_x$ and two branch cuts on the imaginary axis, one from $iq_y$ to $i\infty$ and the other from $-iq_y$ to $-i\infty$.
Using contour integration techniques, we find
\begin{equation}
G(x,q_y)=-i\frac{q_0^2}{q_x}e^{iq_x|x|}-\frac{q_0}{\pi}\int_0^\infty \frac{dt}{\sqrt{t^2+q_y^2}}\frac{t^2}{t^2+q_0^2}\,e^{-\sqrt{t^2+q_y^2}|x|}\,.
\end{equation}
\newline

\noindent\textbf{Normal incidence.}
For normally incident waves
$q_y=0$ and $q_x=q_0$, the incident wave becomes $\phi_i=e^{iq_0x}$, while
the 1D plasmon equation Eq.~\eqref{eqn:plasmon_eq_1d_psi} is reduced to
\begin{equation}
\psi(x)=\frac1{q_0}(G_1\ast V_1) \ast \left[
- \partial_x g(x) \partial_x \phi(x)\right]\,,
\label{eqn:plasmon_eq_1d_psi_normal}
\end{equation}
where $G_1(x)=G(x,0)$ is found to be
\begin{equation}
\begin{split}
G_1(x)&=-iq_0 e^{iq_0|x|}
+\frac{q_0}{\pi}
\left\{\mathrm{Ci}(q_0 |x|)\cos(q_0 |x|)+\left[\mathrm{Si}(q_0 |x|)-\frac{\pi}{2}\right]\sin(q_0 |x|)\right\}\\
&=-iq_0 e^{iq_0|x|}
-\frac{q_0}{2\pi}
\left[e^{iq_0|x|} E_1(iq_0|x|)+e^{-iq_0|x|} E_1(-iq_0|x|)
\right]\,.
\end{split}
\label{eqn:G_1}
\end{equation}
Here $\mathrm{Ci}(z)$, $\mathrm{Si}(z)$ and ${E_1}(z)$ are the cosine, sine and exponential integrals, \begin{equation}
\mathrm{Ci}(z)=-\int_{z}^{\infty}dt\,\frac{\cos{t}}{t},\,
\mathrm{Si}(z)=\int_{0}^{z}dt\,\frac{\sin{t}}{t},\,
E_1(z)=\int_z^\infty dt\,\frac{e^{-t}}{t},
\end{equation}
with the branch cut taken on the negative real axis.
Equation~\eqref{eqn:plasmon_eq_1d_psi_normal} can be further simplified in momentum space in terms of the total electric field  $E=-\partial_x\phi$ and the scattered field $E_s=-\partial_x\psi$,
\begin{equation}
\tilde{E}_s(k) = -ik\tilde{\psi}(k) = \left(\frac{1}{\epsilon(k)}-1\right)(\widetilde{gE})(k)\,,\quad (\widetilde{gE})=\frac1{2\pi}\tilde{g}\ast\tilde{E}\,.
\label{eqn:plasmon_eq_Es_q}
\end{equation}
In the real space this reads,
\begin{equation}
E_s(x) = \int_{-\infty}^{\infty} dx' \left(G_1(x-x')-\delta(x-x')\right)g(x')E(x')\,.
\label{eqn:plasmon_eq_Es_x}
\end{equation}
This is the equation for the scattered field used in the main text.
It indicates  the scattered field $E_s$ can be determined by the conductivity profile $g(x)$ and the local field $E(x)$ around the inhomogeneity using the Green's function $G_1$.

The reflection coefficient $r$ of the normally incident plasmon wave can be obtained by analyzing the scattered field ${E_s}$ at large negative $x$, $E_s\simeq -rE_0e^{-iq_0x}$, where $E_0$ is the amplitude of the incident field $E_i=E_0e^{iq_0x}$.
From Eq.~\eqref{eqn:plasmon_eq_Es_q}, the long range behavior of $E_s$ is determined by the pole of 
$\epsilon^{-1}(k)$ at $k=-q_0$.
Taking the residue, we get
\begin{equation}
r=\frac{iq_0}{E_0}\int_{-\infty}^{\infty}dx \,g(x)E(x)e^{iq_0 x}\,.
\label{eqn:r_E_slit}
\end{equation}
This formula is exact and can be used for any conductivity profile $g(x)$.
The field $E(x)$ can be calculated numerically, cf.~Appendix~\ref{sec:numerical}.

An analytical expression for $E(x)$ (and thus $r$) can be found in the perturbative case, $r\ll 1$,
where the current density is approximately  constant across the inhomogeneity,
\begin{equation}
j=\sigma_0 E_0=\sigma(x)E(x)\,,
\end{equation}
so that $E(x)=\frac{\sigma_0}{\sigma(x)}E_0$ and
\begin{equation}
r=iq_0\int_{-\infty}^{\infty}dx\,\frac{\sigma(x)-\sigma_0}{\sigma(x)} e^{iq_0x}=iq_0\int_{-\infty}^{\infty}dx\,\frac{g(x)}{g(x)+1} e^{iq_0x}\,.
\label{eqn:r_perturb2}
\end{equation}

\section{Reflection coefficient of narrow junctions}

In this section we consider a particular type of conductivity profile,
\begin{equation}
g(x)=g\Theta(a-|x|)\,,\quad g=\frac{\Delta\sigma}{\sigma_0}=\frac{\sigma-\sigma_0}{\sigma_0}\,,
\end{equation}
which describes a junction that has a constant conductivity $\sigma$ and a width $2a$.
Further, we assume the width is narrow compared to the background plasmon wavelength, $a\ll q_0^{-1}$.
In this limit, the leads can be considered perfect metals of effectively infinite conductivity.
This allows us to make simplifications and obtain analytical expressions for the reflection coefficient $r$.
\newline

\noindent\textbf{Vacuum gap.} Analytical solution of the field distribution $E_{\mathrm{vac}}$ for the special case of a vacuum gap, $\sigma=0$ or $g=-1$, is well known,\cite{Smythe1950sde}
\begin{equation}
E_\mathrm{vac}(x) = V F(x)\,, \quad F(x) = \frac{1}{\pi}\frac{\Theta(a-|x|)}{\sqrt{a^2-x^2}}\,,
\label{eqn:F}
\end{equation}
where $V$ is the voltage difference across the junction,
\begin{equation}
V=\int_{-a}^{a}dx\,E(x)\,.
\end{equation}
Substituting $E_\mathrm{vac}(x)$ into Eq.~\eqref{eqn:plasmon_eq_Es_x}, we get
\begin{equation}
\begin{split}
E_i(x)\simeq E_0&=V\int_{-a}^a dx'\,G_1(x-x') F(x)\\
&=V\frac{q_0}{\pi}\left[c +\log\left(\frac{q_0a}{2}\right)-i\pi\right]
\,,\quad -a<x<a\,.
\end{split}
\label{eqn:pre_r_vac}
\end{equation}
Here we used the small-distance approximation for the Green's function,
\begin{equation}
G_1(x)\simeq \frac{q_0}{\pi}\left[c + \log\left(q_0|x|\right)-i\pi\right] + \mathcal{O}(q_0|x|)\,,\quad
q_0|x|\ll1\,,
\end{equation}
where $c\simeq0.577$ is the Euler-Mascheroni constant, and the table integral
\begin{equation}
\int_{-a}^a dx'\,\frac{\log|x-x'|}{{\pi}\sqrt{a^2-x'^2}}=\log\frac{a}2\,.
\label{eqn:convol_G_F}
\end{equation}
(It can be evaluated by the change of variables $x'=a\cos\theta'$.)
Identifying  the capacitance of the junction (cf.~Appendix~\ref{sec:C})
\begin{equation}
C=\frac{\kappa}{2\pi^2}\log\frac{2e^{-c}}{q_0a}\,,
\label{eqn:C2}
\end{equation}
and the reflection coefficient for a narrow junction [Eq.~\eqref{eqn:r_E_slit} with $e^{iq_0x}\to 1$]
\begin{equation}
r\simeq \frac{iq_0}{E_0}gV\,,
\label{eqn:r_narrow}
\end{equation}
our Eq.~\eqref{eqn:pre_r_vac} yields the reflection coefficient of a narrow vacuum gap,
\begin{equation}
r=\frac{i\pi}{\log\left(\frac{2}{q_0a}\right)- c + i\pi}=\frac{i\kappa}{{2\pi C}+i\kappa}\,.
\label{eqn:r_vac2}
\end{equation}
Note that as the gap gets wider, $r$ approaches unity. This can be understood by considering an infinitely wide gap where the current $j(x)$ is completely reflected and is zero at the edge,
so that the current reflection coefficient $r_j=-r=-1$.
\newline

\begin{figure*}[t]
	\centering
	\includegraphics[width=0.95\linewidth]{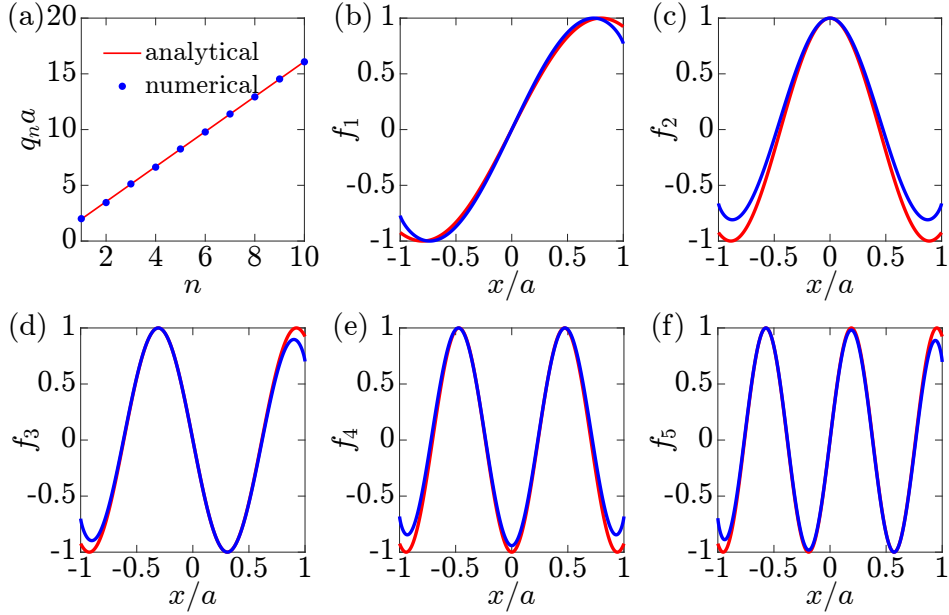}
	\caption{(a) Eigenvalues $q_n$ and (b)-(f) the first five eigenmodes $f_n$ of a junction in a perfect metal sheet.
		The analytical formula is Eq.~\eqref{eqn:eigen_asymp}, while the numerical results are obtained by solving Eq.~\eqref{eqn:E_n_eigen_eq}.}
	\label{fig:eigen}
\end{figure*}

\noindent\textbf{General cases.} We 
add $E_i$ to both sides of Eq.~\eqref{eqn:plasmon_eq_Es_x} to get
\begin{equation}
E_i(x)\simeq E_0 = -\left[G(x)\ast gE(x)\right] + (1+g)E(x)\,,\quad -a<x<a\,.
\label{eqn:E_E0}
\end{equation}
We multiply this by $\sigma_0 F(x)$, where $F(x)$ is given by Eq.~\eqref{eqn:F}, and integrate it from $x=-a$ to $a$,
\begin{equation}
\sigma_0 E_0 = -\sigma_0 g V \frac{q_0}{\pi}\left[c + \log\left(\frac{q_0a}{2}\right)-i\pi\right]
+\sigma_0(1+g)\int_{-a}^{a}dx E(x) F(x)\,.
\label{eqn:pre_Shockley_Ramo}
\end{equation}
We arrive at the equation (mentioned in the main text)
\begin{equation}
\sigma_0 E_0 (1 - r) = -i\omega V C'+\int_{-a}^{a}dx\, \sigma E(x) F(x)\,, \quad C' \equiv -g C\,,
\end{equation}
which represents the conservation of charge.
Note that this equation has the correct limiting behavior, reproducing the reflection coefficient for the vacuum gap when $g=-1$ and $\sigma=0$, and yielding $r=0$ when $g=0$ and $\sigma=\sigma_0$.

Equation~\eqref{eqn:pre_Shockley_Ramo} can be used to find the reflection coefficient $r$ if the field $E(x)$ were known.
We use the following ansatz for the field,
\begin{equation}
E(x)=d_0+\sum_{n=1}^{\infty} d_n E_n(x)\,,\quad -a<x<a\,,
\label{eqn:E_series}
\end{equation}
which is an expansion into the eigenmodes $E_n$ of the system when there is no external field.
The eigenmodes $E_n$ have the following properties,
\begin{equation}
E_n(|x|>a)=0\,,\quad V_n =\int_{-a}^a dx\, E_n(x)=0\,,
\label{eqn:E_n_properties}
\end{equation}
as the system now consists of a junction with nonzero conductivity $\sigma_n$ and two sheets of infinite conductivity.
The eigenvalue equation governing $E_n(x)$ can be derived 
from Eq.~\eqref{eqn:E_E0} using the substitutions $g=-1$ and $1+g=\sigma_n/\sigma_0$ and invoking Eqs.~\eqref{eqn:convol_G_F} and \eqref{eqn:E_n_properties},
\begin{equation}
E_n(x)=A-\frac{q_n}{\pi}\int_{-\infty}^{\infty}dx'\, \log\frac{|x-x'|}{L}\,  E_n(x')\,,
\label{eqn:E_n_eigen_eq}
\end{equation}
where $q_n=i\kappa\omega/2\pi\sigma_n$,
the constant $L$ is arbitrary, and the constant $A$ can be obtained by multiplying both sides of Eq.~\eqref{eqn:E_n_eigen_eq} by $F(x)$ and integrating over $x$,
\begin{equation}
A=\int_{-a}^{a} E_n(x) F(x)\,.
\end{equation}
Since the kernel in Eq.~\eqref{eqn:E_n_eigen_eq} is self-adjoint, the eigenmodes $E_n$ are orthogonal,
\begin{equation}
\int_{-a}^a dx\,E_n(x) E_m(x)=0\,,\quad n\neq m\,,
\end{equation}
and the eigenvalues $q_n$ are real.
Eq.~\eqref{eqn:E_n_eigen_eq} can be solved numerically to obtain the eigenmodes.
As shown in Fig.~\ref{fig:eigen}, they have the asymptotic forms
\begin{equation}
\frac{E_n}{E_0}\simeq\cos\left(q_n x -\frac{n\pi}2\right)\,,\quad q_n\simeq\frac{\pi}{2a}\left(n+\frac14\right)\,,
\label{eqn:eigen_asymp}
\end{equation}
as $n$ is increased.

Next, we determine the coefficients $d_n$ for the fields $E_n$. 
Under an external field the voltage across the junction $V\neq0$ but all $V_n=0$.
Thus $V$ must be accounted for by $d_0$, i.e.,
\begin{equation}
d_0=\frac{V}{2a}\,.
\end{equation}
To find $d_n$, we rewrite Eq.~\eqref{eqn:E_E0} using Eqs.~\eqref{eqn:E_series} and \eqref{eqn:E_n_eigen_eq} as
\begin{equation}
\log\frac{|x|}{L}\ast h(x)=0\,,
\end{equation}
with a solution
\begin{equation}
h(x)=V\left[\frac1{2a} - F(x)\right] + \sum_{n=1}^{\infty}\left(1+\frac{g+1}{g}\frac{q_n}{q_0}\right)d_nE_n=0\,.
\end{equation}
Here Eq.~\eqref{eqn:convol_G_F} was used to convert constants into the convolution,
\begin{equation}
1 = \frac{1}{\log(a/2)}\, F(x) \ast \log\frac{|x|}{L}\,,
\end{equation}
and 
Eq.~\eqref{eqn:pre_Shockley_Ramo} was used in the form
\begin{equation}
E_0-(g+1)\int_{-a}^{a} dx\, E(x) F(x) = -g \frac{q_0}{\pi}\, V \log\frac{a}2\,.
\end{equation}
Using the orthogonality of the eigenmodes, we find
\begin{equation}
d_n=\frac{V}{1+\frac{g+1}{g}\frac{q_n}{q_0}}\frac{\int_{-a}^{a} dx\, E_n(x) F(x)}{\int_{-a}^{a}dx\,E_n^2(x)}\,.
\end{equation}
Note that in the limit of a vacuum gap where $g=-1$, $d_n$ reduces to coefficients of the expansion of $E_\mathrm{vac} = V F(x)$ in the $E_n$ basis, as expected.
In the manuscript, the dimensionless expansion coefficients $b_0$ and $b_n$ are defined as
\begin{equation}
b_0=\frac{d_0}{(V/2a)} = 1\,,\quad b_n=d_n\frac{E_0}{(V/2a)} = 
\frac{2a}{1+\frac{g+1}{g}\frac{q_n}{q_0}}\,
\frac{\int_{-a}^{a} dx\,f_n(x) F(x)}{\int_{-a}^{a}dx\,f_n^2(x)}\,,
\end{equation}
and the field is expanded as 
\begin{equation}
E(x)=\frac{V}{2a}\left(1+\sum_{n=1}^\infty b_n f_n\right)\,,
\qquad f_n = \frac{E_n}{E_0}\,.
\end{equation}

\noindent\textbf{Analytic approximation.}
Having found the coefficients $d_n$, the expression for the reflection coefficient can now be written from Eq.~\eqref{eqn:pre_Shockley_Ramo} using Eq.~\eqref{eqn:r_narrow},
\begin{equation}
r^{-1} = 1 - \frac{2\pi i}\kappa C - \frac{i}{q_0} \frac{g+1}{g}
\left\{
\frac1{2a}
+ \sum_{n=1}^{\infty}\frac{1}{1+\frac{g+1}{g}\frac{q_n}{q_0}}\,
\frac{\left[\int_{-a}^{a} dx\,f_n(x) F(x)\right]^2}{\int_{-a}^{a}dx\,f_n^2(x)}
\right\}\,.
\end{equation}
An analytical expression for $r$ can be found using the approximation Eq.~\eqref{eqn:eigen_asymp}, so that
\begin{equation}
\int_{-a}^{a}dx\,f_n^2(x) = a\left[1 + \frac1{\sqrt{2}\pi\left(n+\frac14\right)}\right]\,,
\end{equation}
and
\begin{equation}
\begin{split}
\left[\int_{-a}^{a}dx\,f_n(x) F(x)\right]^2 &= 0\,,\quad n\ \mathrm{odd}\\
&= J_0^2(q_n a) \simeq\frac{2+\sqrt{2}}{\pi^2}\frac{1}{n+\frac14}\,,\quad n\gg 1\ \mathrm{even}\,,
\end{split}
\end{equation}
where $J_0(z)$ is the Bessel function of the first kind.
The summation is then
\begin{equation}
\sum_{n=1}^{\infty}\frac{1}{1+\frac{g+1}{g}\frac{q_n}{q_0}}\frac{\left[\int_{-a}^{a} dx\,f_n(x) F(x)\right]^2}{\int_{-a}^{a}dx\,f_n^2(x)}\simeq
\frac{2q_0}{\pi}\frac{g}{g+1}\frac{2+\sqrt{2}}{\pi^2}\sum_{n=1}^\infty\frac{1}{(2n+\alpha)(2n+\beta)}\,,
\end{equation}
where $\alpha=\frac{2q_0a}{\pi}\frac{g}{g+1}+\frac14$ and $\beta=\frac1{\sqrt{2}\pi}+\frac14$.
Using the identity
\begin{equation}
\sum_{n=1}^\infty\frac{1}{(2n+\alpha)(2n+\beta)}
= \frac{\Psi\left(1+\frac{\alpha}{2}\right)-\Psi\left(1+\frac{\beta}{2}\right)}{2(\alpha-\beta)}\,,
\end{equation}
where $\Psi(z)=\Gamma'(z)/\Gamma(z)$ is the digamma function, the reflection coefficient is 
\begin{equation}
\begin{split}
r^{-1}\simeq 1 - & \frac{2\pi i}\kappa C - \frac{i}{2q_0a}\frac{g+1}{g}\\
&-i\frac{\sqrt{2}+1}{\pi^2}\frac{2(g+1)}{2\sqrt{2}q_0ag-(g+1)}\left[
\Psi\left(\frac98+\frac{q_0a}{\pi}\frac{g}{g+1}\right)
- \Psi\left(\frac98+\frac{\sqrt{2}}{4\pi}\right)\right],
\end{split}
\label{eqn:r_approx_analyt}
\end{equation}
in the limits $q_0a\ll 1$ and $n\gg 1$.

\section{Capacitance of a vacuum gap}
\label{sec:C}

The capacitance Eq.~\eqref{eqn:C2} found in the previous section can be simply derived by considering a vacuum gap in a perfect metal sheet.
The induced charge distribution on the sheet is well-known,\cite{Smythe1950sde}
\begin{equation}
\delta\rho(x)=-\frac{\kappa V}{2\pi^2}\frac{1}{\sqrt{x^2-a^2}}\,\Theta(|x|-a)\,\mathrm{sgn}(x)\,.
\label{eqn:rho_perfect}
\end{equation}
The capacitance $C$ is then
\begin{equation}
C=\frac{Q}{V} = \frac1V\int\limits_{-L}^{-a}dx\,\delta\rho(x)=\frac{\kappa}{2\pi^2}\log\left(\frac{2L}{a}\right)\,,
\end{equation}
where $L$ is the long-distance cutoff length.
If $\sigma_0$ is finite, the charge distribution $\delta\rho(x)$ evolves into plasmonic waves at distance $\sim q_0^{-1}$.
Therefore, we expect $L \sim q_0^{-1}$.
From the expression for $C$ in the previous section we get the exact relation $L=e^{-c}q_0^{-1}$.

\section{Numerical calculation of the reflection coefficient}
\label{sec:numerical}

The reflection coefficient of a conductivity profile $g(x)$ of any shape and width can be calculated numerically by recasting Eq.~\eqref{eqn:plasmon_eq_Es_x}
in the following form,
\begin{equation}
E=E_s+E_i=G\ast(gE)-gE+E_i\,,
\end{equation}
or
\begin{equation}
E_i=\mathcal{K}E\,,\quad \mathcal{K}=1+g-G\ast g\,.
\end{equation}
Inversion of $\mathcal{K}$ then yields the total field within the inhomogeneity $E=\mathcal{K}^{-1}E_i$ given the incident field $E_i$, 
and the reflection coefficient can be found using Eq.~\eqref{eqn:r_E_slit}.
The kernel $\mathcal{K}$ has elements
\begin{equation}
\mathcal{K}_{ij}=\left(1+g(x_i)\right)\delta_{ij}-G(x_i-x_j) g(x_j)\Delta x_j\,,
\end{equation}
where $x$ within the junction is discretized into $x_i$ and 
the Green's function $G$ is calculated using Eq.~\eqref{eqn:G_1}.
For our calculation we chose the discretization
\begin{equation}
x_i=a\cos \frac{\pi}{N}\left(i-\frac12\right) \,,\quad i=N,N-1,...,1\,,
\end{equation}
and
\begin{equation}
\Delta x_i=a\frac{\pi}{N}\sin \frac{\pi}{N}\left(i-\frac12\right) \,,
\end{equation}
with a number of grid points $N\sim 10^2$.
The diverging diagonal elements $G(x_i-x_i)$ is regularized using $G(\Delta x_i/2\pi)$, explained as follows.
For small arguments, $G(x)$ is proportional to $\log|x|$.
Assume we wish to calculate the integral $\int_{-L}^{L} dx\,\log|x|$ as a discrete sum on a uniform grid.
The grid is taken to be $x_i=i\Delta x$ with $i=-N,...,N$, which has $2N+1$ points separated by distance $\Delta x$ and a total length $2L=(2N+1)\Delta x$.
Writing $L_0$ as the regularized replacement for the diverging $\log0$, we have
\begin{equation}
\int_{-L}^{L} dx\,\log|x|=\left(L_0+\sum_{i\neq 0}\log|x_i|\right)\Delta x\,.
\end{equation}
The summation can be simplified using Stirling's approximation for large $N$,
\begin{equation}
\sum_{i=1}^{N}\log|x_i|=\log\left(\Delta x^N N!\right)\simeq\log\left(\Delta x^N\right)+\log\left[\sqrt{2\pi N}\left(\frac{N}{e}\right)^N\right]\,.
\end{equation}
After some algebra we find $L_0=\log(\Delta x/2\pi)$, prescribing the regularization $G(\Delta x_i/2\pi)$. 

\begin{figure}[t]
	\centering
	\includegraphics[width=0.7\linewidth]{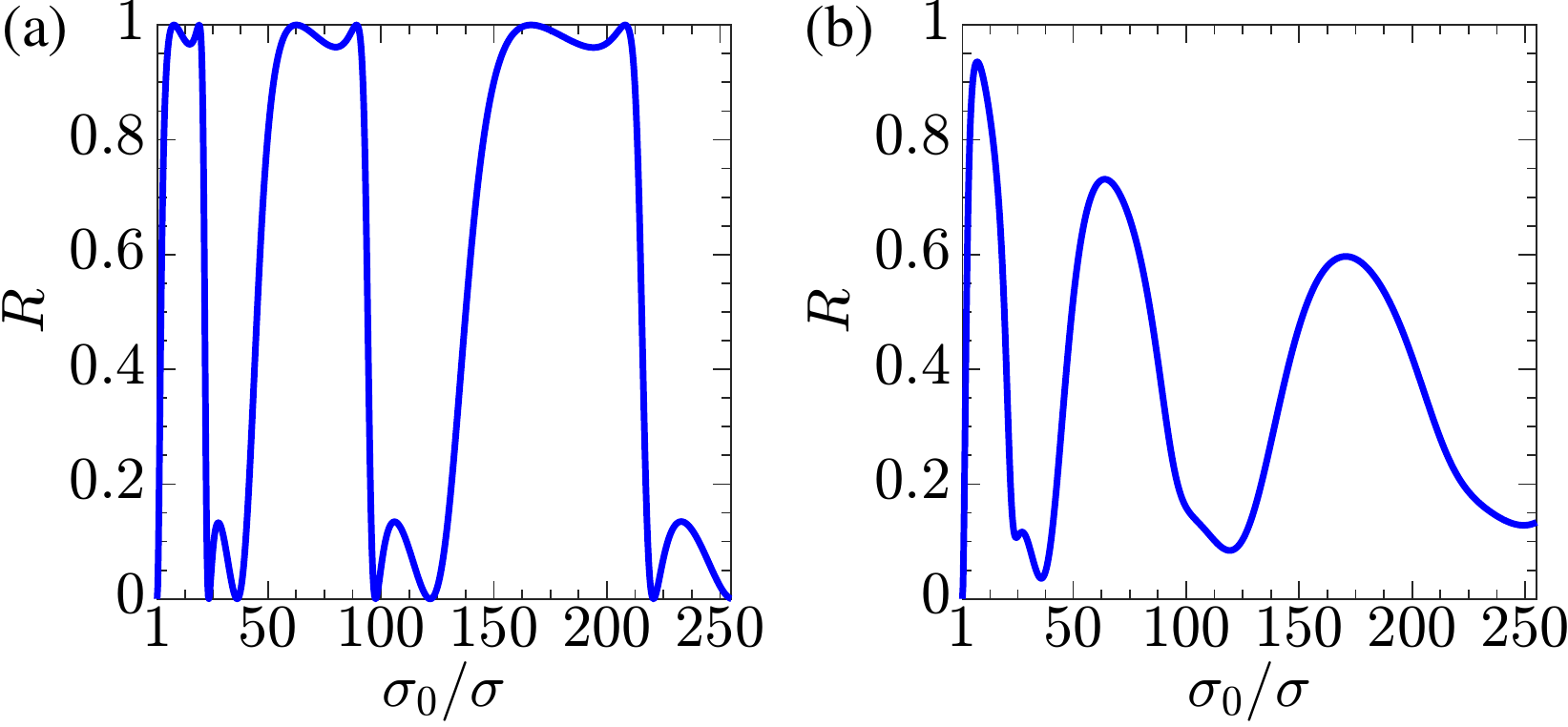}
	\llap{\raisebox{0.22\linewidth}{\includegraphics[height=0.08\linewidth]{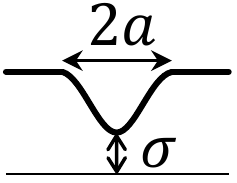}\hspace{0.04\linewidth}}}
	\caption{
		(a) Reflectance of a junction with a smooth conductivity profile for $a=0.1\lambda_0$ and $\gamma=0$.
		(b) Similar plot at $\gamma=0.05$.
		(Inset) The conductivity profile $\sigma_\mathrm{sm}$.
	}
	\label{fig_smooth}
\end{figure}

\section{Smooth conductivity profiles}
In our model we assume
sharp conductivity changes at the edges of the junction.
This may be difficult to achieve in experiment, or yield unphysical results in theory.
Here we show that the physics for a sharply-changing conductivity profile remains qualitatively the same for a smoothly-varying one.
To demonstrate, we calculated numerically the reflectance of a smooth conductivity profile 
\begin{equation}
\sigma_\mathrm{sm}(|x|<a)=\sin^2\left(\frac{\pi}{2a}|x|\right)+\frac{\sigma}{\sigma_0}\cos^2\left(\frac{\pi}{2a}|x|\right)\,,
\end{equation} 
where $\sigma$ now denotes the lowest value of conductivity occuring at $x=0$.
As shown in Fig.~\ref{fig_smooth}, the reflectance retains the same features -- capacitive open-circuit resonance and Fano-shaped cavity resonances with weak odd modes.
The open-circuit resonance is still persistent under damping, while
the cavity resonances are less sharp due to a larger $|t'|$ for smooth profiles.
Note that to get an anti-resonance at the same location, $\sigma_0/\sigma\approx 10$, the junction width had to be increased by an order of magnitude, from $a/\lambda_0=0.01$ to $0.1$.

\section{Phase of the reflection coefficient}

\begin{figure}[t]
	\centering
	\includegraphics[width=0.7\linewidth]{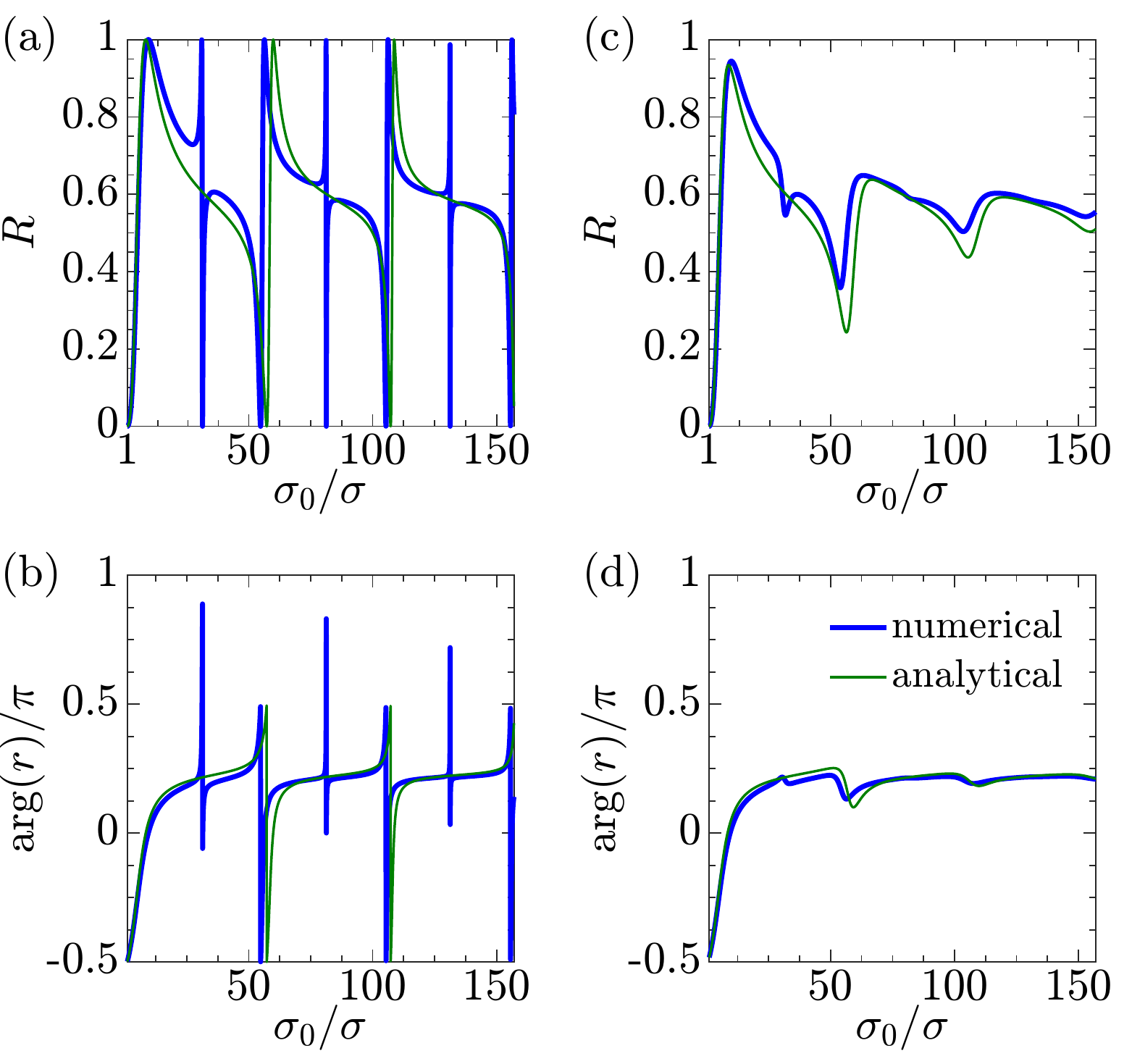}
	\caption{
		(a) Squared magnitude $R=|r|^2$ and (b) phase $\mathrm{arg}(r)$ of the reflection coefficient $r$ for a junction of width $a=0.01\lambda_0$ and damping $\gamma=0$. 
		(c, d) Similar quantities for $\gamma=0.05$.
		The blue curves are calculated numerically as described in Appendix~\ref{sec:numerical}, while the green curves are from Eq.~\eqref{eqn:r_approx_analyt}.
	}
	\label{fig_r_theta}
\end{figure}

We show in Fig.~\ref{fig_r_theta} the phase of the reflection coefficient for a narrow junction $a=0.01\lambda_0$.
In the perturbative regime where $\sigma\lesssim \sigma_0$ the phase is $-\pi/2$, in agreement with Eq.~\eqref{eqn:r_perturb2}.
At the open-circuit resonance the phase is $0$, $r>0$ is real and $|r|\simeq 1$. 
For large conductivity contrasts the phase approaches the limiting value $\simeq0.24\pi$ predicted by Eq.~\eqref{eqn:r_vac2}.

\section{Effect of damping contrast}
\label{sec:damping}

\begin{figure}[t]
	\centering
	\includegraphics[width=0.35\linewidth]{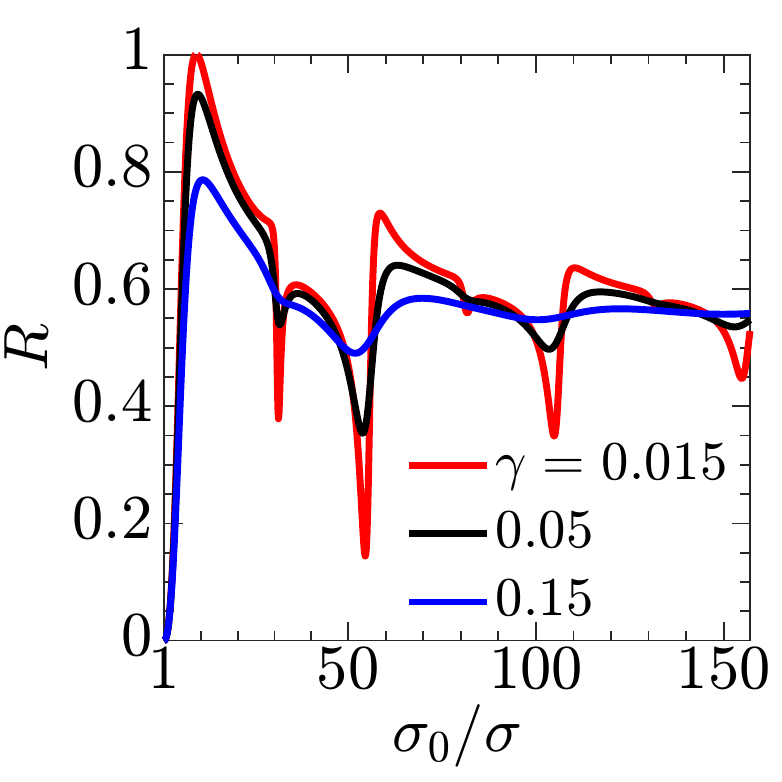}
	\caption{
		Reflectance of the junction when it has a damping $\gamma$ different from or equal to the background value $\gamma_0=0.05$.
		Junction width $a=0.01\lambda_0$.
	}
	\label{fig_g_complex}
\end{figure}

In the manuscript we only considered  the case when the damping $\gamma$ in the junction is the same  as in the sheet, $\gamma=\gamma_0$.
One might expect a stronger damping in the junction would further dampen the resonances and vice versa.
This is indeed the case, as can be seen from the reflectances calculated under different values of $\gamma$ shown in Fig.~\ref{fig_g_complex}.
The conductivities  are parametrized as
\begin{equation}
\sigma_0=\frac{S_0}{1+i\gamma_0}\,,\quad \sigma=\frac{S}{1+i\gamma}\,,
\end{equation}
where $\gamma_0=0.05$ and $\gamma$ takes the values $0.015$, $0.05$, and $0.15$.
The resonant features are enhanced if $\gamma<\gamma_0$ and diminished if $\gamma>\gamma_0$.

\section{Comments on previous works}
\label{sec:prev_work}

The reflection from wide junctions has been widely studied \cite{Dyer2012iit,Dyer2013itc,Sydoruk2015trt}.
These works either did not take into account the nontrivial phase shift of the reflection coefficient, or limited themselves to approximated and numerical results. The analytical form for the phase shift was first reported in Ref. \citenum{Rejaei2015ssp}.

The reflection from a narrow junction was previously studied in Ref. \citenum{Ryzhii2003asd}, where the electrostatic potential
within the junction was incorrectly assumed to be
\begin{equation}
\phi(x, z)=\frac12 V \frac{\sin qx}{\sin qa}e^{-q|z|}\,,\quad |x|\leq a\,.
\label{eqn:Ryzhii_ansatz}
\end{equation}
At first glance, this solution seems to satisfy Laplace's equation 
\begin{equation}
\nabla^2 \phi = \frac{4\pi}{\kappa}\delta\rho(x) \delta(z)
\end{equation}
within the junction under the boundary conditions
\begin{equation}
\phi(x\geq a,0)=\frac12 V\,,\quad \phi(x\leq -a,0)=-\frac12 V\,.
\label{eqn:Ryzhii_bc}
\end{equation}
However, Eq.~\eqref{eqn:Ryzhii_ansatz} does not satisfy the boundary conditions \textit{outside} the junction.
If Eq.~\eqref{eqn:Ryzhii_ansatz} were true, the potential on the plane $x=a$ would be 
\begin{equation}
\phi(a,z)=\frac{V}2e^{-q|z|}\,.
\end{equation}
Using the well-known result for Laplace's equation, \cite{Morse_book}
\begin{equation}
\phi(x,z)=\frac{\tilde{x}}{\pi}\int_{-\infty}^{\infty}dz'\,\frac{\phi(a,z')}{(z'-z)^2+\tilde{x}^2}
\end{equation}
where $\tilde{x}\equiv x-a$, the potential on the $z=0$ plane would be
\begin{equation}
\phi(x>a,0)=\frac{V}{2\pi}\tilde{x}\int_{-\infty}^{\infty}dz'\,\frac{e^{-q|z'|}}{z'^2+\tilde{x}^2}\,.
\label{eqn:Ryzhii_phi_lead}
\end{equation}
Eq.~\eqref{eqn:Ryzhii_phi_lead} yields $\phi(x\to\infty,0)=0\neq \frac{V}2$, so the boundary condition Eq.~\eqref{eqn:Ryzhii_bc} is \textit{not} satisfied.
The failure of the ansatz Eq.~\eqref{eqn:Ryzhii_ansatz} can be simply understood as follows.
At the interface between two half-planes with different plasmon momenta $q_1$ and $q_2$, the $z$ dependence of the potential must change from $e^{-q_1|z|}$ to $e^{-q_2|z|}$ across the interface.
Therefore a solution with a fixed $e^{-q|z|}$ across the entire junction cannot be correct.

\bigskip \noindent \textbf{Funding}\\ \\
The work at UCSD in its applications to superconductors is supported by the Department of Energy under Grant DE-SC0012592.
Applications to graphene are supported by the Office of Naval Research under Grant N00014-15-1-2671 and those to semiconductors by the National Science
Foundation under Grant ECCS-1640173 and by the Semiconductor Research Corporation (SRC) through the Center for Excitonic Devices at University of California, San Diego, research 2701.002. EJM is supported by the Department of Energy through grant DE FG02 84ER45118.

\bibliography{nanogap}
\end{document}